\documentclass[journal]{IEEEtran}

\usepackage{amsmath,amsfonts,amssymb,amsthm}

\usepackage{graphicx}
\usepackage{subfig} 
\usepackage{booktabs} 
\usepackage{multirow}
\usepackage{threeparttable} 
\usepackage{array} 

\usepackage[linesnumbered,boxed,ruled,commentsnumbered]{algorithm2e}
\usepackage{stfloats} 
\usepackage{textcomp} 
\usepackage{url} 
\usepackage{cite} 
\usepackage{soul}
\usepackage{xcolor}
\usepackage{bm}
\usepackage{autobreak}

\begin{document}
\title{Bayesian Critique-Tune-Based Reinforcement Learning with Adaptive Pressure for Multi-Intersection Traffic Signal Control}
\author{Wenchang Duan, ~\IEEEmembership{Graduate Student Member,~IEEE,}
Zhenguo Gao, 
Jiwan He, ~\IEEEmembership{Graduate Student Member,~IEEE,}
Jinguo Xian, ~\IEEEmembership{Member,~IEEE}
%<-this % stops a space
% ~\IEEEmembership{Graduate Student Member,~IEEE,}}
%         John~Doe,~\IEEEmembership{Fellow,~OSA,}
%         and~Jane~Doe,~\IEEEmembership{Life~Fellow,~IEEE}% <-this % stops a space
\thanks{This work was supported in part by the National Natural Science Foundation of China under Grant 12001365. 

Wencang Duan, Zhenguo Gao and Jinguo Xian are with the School of Mathematical Sciences, Shanghai Jiao Tong University (SJTU), Shanghai 200240, China (e-mail: duanwenchang@sjtu.edu.cn, gaozheng@sjtu.edu.cn, kih020429@sjtu.edu.cn, jgxian@sjtu.edu.cn). Correspondence to: Zhenguo Gao.}}
\maketitle

\begin{abstract}
Adaptive Traffic Signal Control (ATSC) system is a critical component of intelligent transportation, with the capability to significantly alleviate urban traffic congestion. Although reinforcement learning (RL)-based methods have demonstrated promising performance in achieving ATSC, existing methods are still prone to making unreasonable policies. Therefore, this paper proposes a novel Bayesian Critique-Tune-Based Reinforcement Learning with Adaptive Pressure for multi-intersection signal control (BCT-APLight). In BCT-APLight, the Critique-Tune (CT) framework, a two-layer Bayesian structure is designed to refine the excessive trust of RL policies. Specifically, the Bayesian inference-based Critique Layer provides effective evaluations of the credibility of policies; the Bayesian decision-based Tune Layer fine-tunes policies by minimizing the posterior risks when the evaluations are negative. Meanwhile, an attention-based Adaptive Pressure (AP) mechanism is designed to effectively weight the vehicle queues in each lane, thereby enhancing the rationality of traffic movement representation within the network. Equipped with the CT framework and AP mechanism, BCT-APLight effectively enhances the reasonableness of RL policies. Extensive experiments conducted with a simulator across a range of intersection layouts demonstrate that BCT-APLight is superior to other state-of-the-art (SOTA) methods on seven real-world datasets. Specifically, BCT-APLight decreases average queue length by \textbf{\(\boldsymbol{9.60\%}\)} and average waiting time by \textbf{\(\boldsymbol{15.28\%}\)}. Codes are open-sourced. 
\end{abstract}

% \fi
\IEEEpeerreviewmaketitle

\section{Introduction}

\IEEEPARstart{W}{ITH} urban populations continuous growth and cities expanding, traffic congestion has become increasingly severe, placing escalating pressure on the environment and economy \cite{r1,r2,r3}. As a significant component of transportation systems, ATSC system can effectively alleviate traffic congestion\cite{r4,r5,r6}. Reinforcement learning (RL) has been extensively explored as an efficient method in the ATSC system \cite{r9,r10,r11,r12}. However, the increasingly complex traffic demands of modern cities have exceeded the capabilities of existing RL-based methods \cite{r13,r14,r15,r16,r17}, frequently leading to unreasonable policies. Therefore, it is crucial to enhance the reasonableness of RL policies.

By effectively optimizing long-term returns and enabling dynamic interactions with environments, RL has demonstrated promising performance \cite{r18,r19,r20}. Specifically, references \cite{r21}\cite{r22} employed the independently RL agent at each intersection to improve the scalability issues. References \cite{r23}\cite{r24} introduced graph networks into RL, which implemented parameter sharing mechanisms to incorporate temporal and spatial influences from neighboring intersections. Da et al. \cite{r25} proposed the PromptGAT method to bridge the simulation-to-reality performance gap. However, the above methods overlook the reasonableness of RL policies.

To address the unreasonableness of the RL policies, much literature has focused on how to enhance the reliability of the learning process \cite{r27}. Specifically, Wang et al. proposed a cooperative double Deep Q-Network (DQN) method \cite{r28} to improve the robustness of the learning process. Hinton et al. proposed a teacher-student framework \cite{r29}, in which the teacher module guides the student module to avoid major errors in the learning process. And references \cite{r30,r31} addressed the limitation of traditional teacher-student advising methods, which only offered advice in situating the same state or after having similar experiences. Furthermore, references \cite{r32,r33,r34} introduced the A2C algorithm, allowing for the evaluation and adjustment of the learning process. References \cite{r35,r36} proposed the multi-objective Bayesian optimization to solve the model-free problem in the learning process. However, the aforementioned methods simply focus on the learning process while overlooking the policy-making process, which often leads to excessive trust of RL policies.

In the ATSC system, effective traffic movement representation is also crucial to enhance the reasonableness of RL policies \cite{r9}\cite{r10}. To efficiently represent the traffic scenario, references \cite{r40, r41, r56} developed methods to quantify complex traffic information based on the max-pressure theory. Wu et al. \cite{r42} introduced the concept of efficient pressure to represent traffic movement, achieving notable efficiency in signal control. Zhang et al. \cite{r53} further incorporated both waiting and running vehicles to enhance the comprehensiveness of traffic movement representation. However, the existing pressure calculation methods overlook the varying influence of multiple upstream lanes on each downstream lane. This limitation may amplify the impact of low-traffic lanes while reducing responsiveness to high-traffic ones, thereby leading to ineffectiveness of traffic movement representation.

\begin{figure*}[t]
    \centering
    \makebox[\textwidth][c]{ 
        \includegraphics[width=1.03\textwidth]{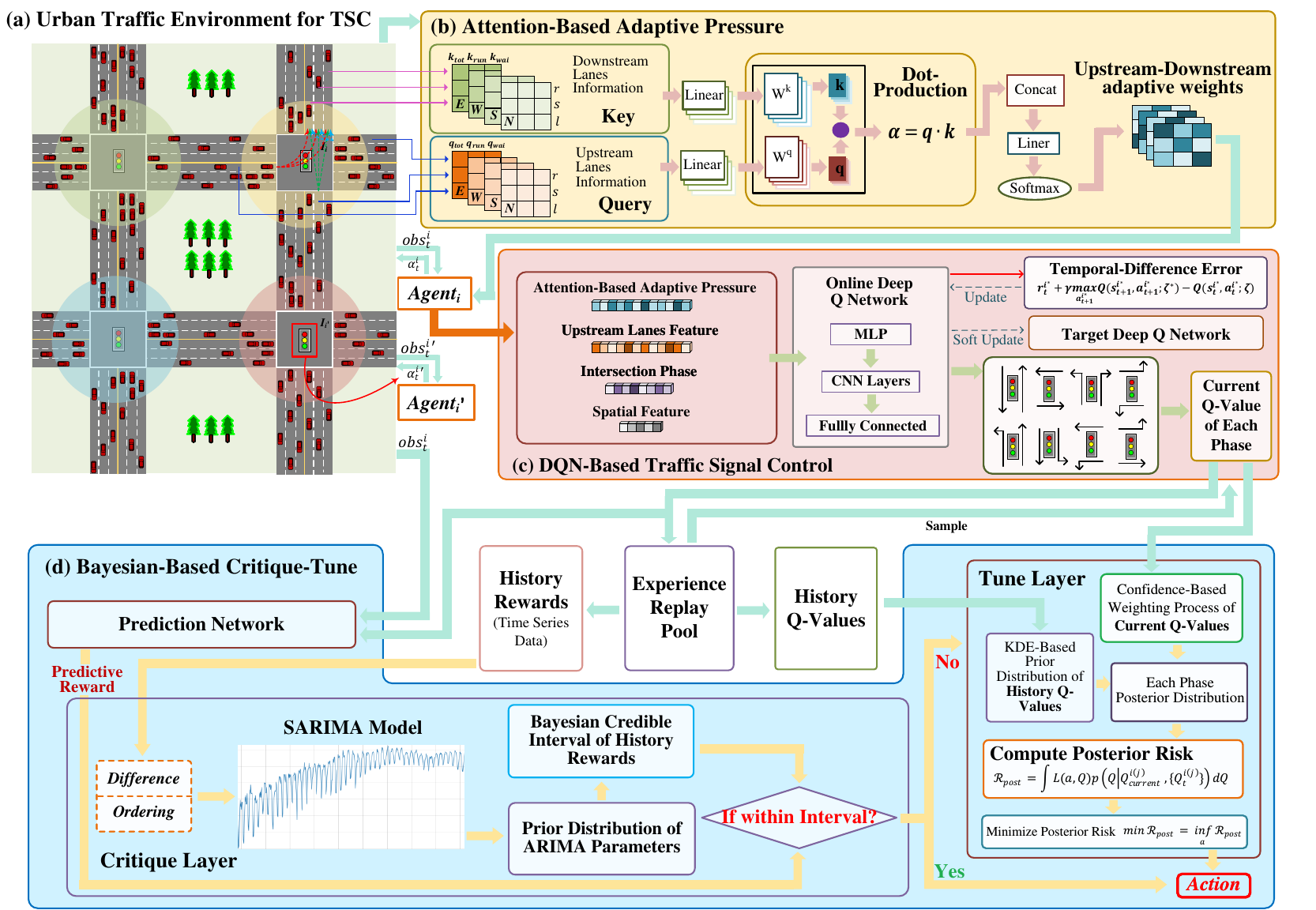} 
    }
    \caption{Architecture of BCT-APLight. The urban traffic environment for ATSC (\textbf{a}) offers complex traffic dynamics and an interactive framework for reinforcement learning (RL). Attention-based Adaptive Pressure (\textbf{b}) enables RL agents to effectively capture traffic features. This adaptive pressure, combined with traffic lane details, enhances DQN-based traffic signal control (\textbf{c}). The Bayesian-based Critique-Tune framework (\textbf{d}) evaluates and refines RL policies for improved decision-making.}
    \label{fig:1}
\end{figure*}

According to the aforementioned analysis, the excessive trust of RL policies and the ineffectiveness of traffic movement representation severely limit the formulation of reasonable policies. To address these problems, this paper proposes a Bayesian Critique-Tune-Based Reinforcement Learning with Adaptive Pressure for multi-intersection signal control (BCT-APLight). The framework of the BCT-APLight is shown in Fig. \ref{fig:1}. Firstly, the Critique-Tune (CT) framework employs a two-layer Bayesian structure to refine RL policies. Specifically, the Bayesian inference-based Critique Layer constructs a Bayesian credible interval, which using historical rewards to evaluate current policies. If the evaluation is negative, the Bayesian decision-based Tune Layer calculates the posterior risk of each phase according to the posterior probability of Q-values. By minimizing this risk, the Tune Layer fine-tunes the policies. In addition, an attention-based Adaptive Pressure (AP) is designed for measuring the vehicle queues from each upstream to each downstream. By dynamically updating the weight, the AP mechanism achieves effective traffic movement representation for ATSC. With the above designs, the BCT-APLight can effectively enhancing the reasonableness of RL policies.

The contributions of this paper can be summarized as follows.
\begin{itemize}
    \item This paper proposes a Bayesian Critique-Tune-based Reinforcement Learning with Adaptive Pressure for traffic signal control. Equipped with a Critique-Tune framework, BCT-APLight achieves to refine the excessive trust of RL policies.
    \item An attention-based Adaptive Pressure is proposed to effectively capture the traffic features. Thereby effectively enhancing the rationality of traffic movement representation in real time.
    \item This paper conducts extensive experiments on real-world traffic datasets and compare BCT-APLight with existing methods. The experimental results demonstrate that BCT-APLight decreases average queue length by \(9.60\%\) and average waiting time by \(15.28\%\) on average compared with Advanced-CoLight. %Codes are open sourced in https://github.com/duanwenchang/BCT-APLight. 
\end{itemize}

The rest of this paper is organized as follows. Section \ref{sectionII} describes the ATSC in urban intersections and models ATSC problem based on Markov Decision Process (MDP). Section \ref{sectionIII} presents Bayesian Critique-Tune framework for RL. Section \ref{sectionIV} presents the RL with attention-based Adaptive Pressure and its training process. Section \ref{sectionV} gives the performance of the proposed method. Section \ref{sectionVI} draws conclusions.

\section{Traffic Signal Control with Reinforcement Learning in Urban Intersections}
\label{sectionII}

This section describes the concepts and methodology relevant to this paper. Firstly, key definitions related to ATSC are provided in Subsection \ref{sec:subsecIIA}. Following that, Subsection \ref{sec:subsecIIB} introduces the mathematical framework of reinforcement learning for ATSC. Based on these concepts and methodology, this paper researches adaptive traffic signal control across multi-intersections to minimize average travel time.

\begin{figure}[!t]
    \centering
    \hspace*{-0.7cm}
    \subfloat[Illustration of AP]{%
        \includegraphics[width=0.9\columnwidth]{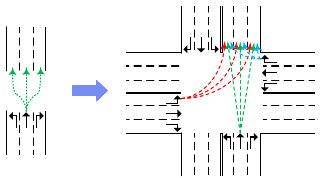}} \\
    \vspace{0.5cm}
    \hspace*{-0.7cm}
    \subfloat[Traffic signal phase]{%
        \includegraphics[width=0.7\columnwidth]{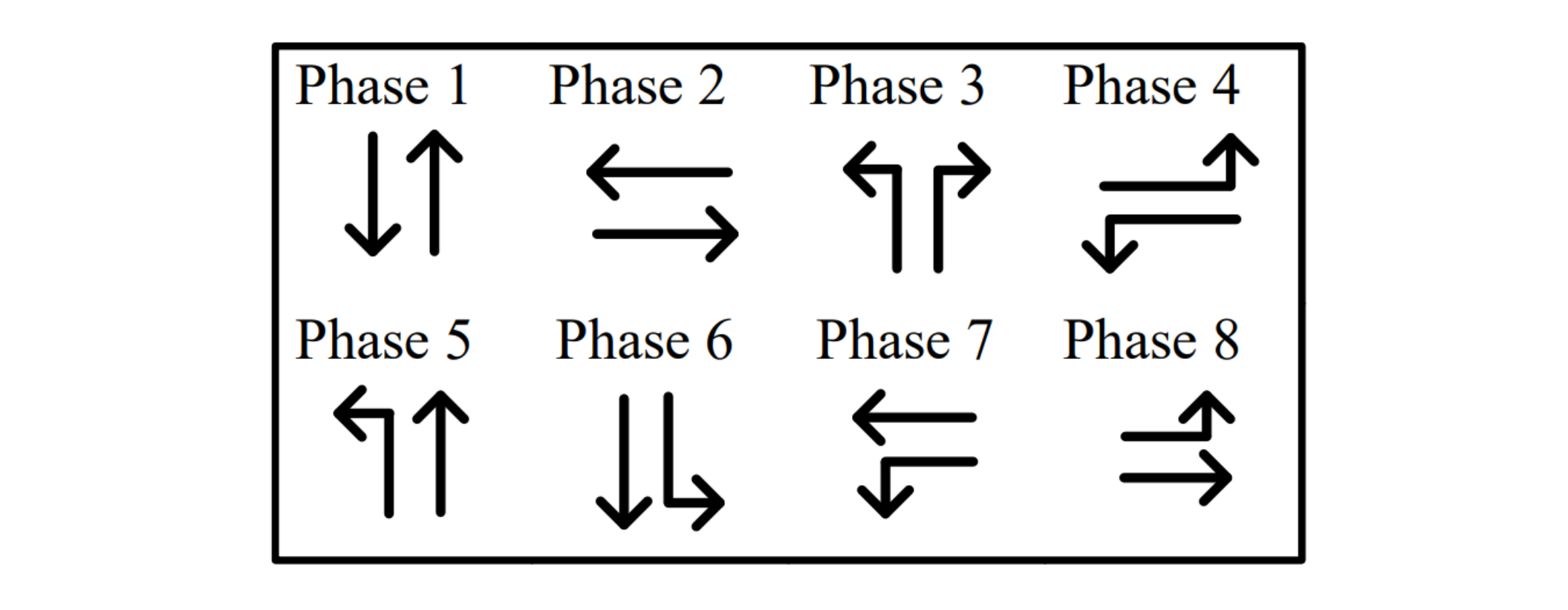}}
    \caption{The traditional efficient pressure and the attention-based adaptive pressure are provided in (a). The eight traffic signals are illustrated in (b).}
    \label{fig:2}
\end{figure}

\subsection{ATSC Definition in Multi-Intersections}
\label{sec:subsecIIA}

\textit{Definition 1 (Traffic Intersection and Road):} The traffic network can be modeled as a directed graph, where nodes correspond to \textit{n} number of intersections \textit{I} and edges represent roads. At each intersection \textit{$I_i$}, the road network consists of four directions \textit{\{E,W,S,N\}}: east (\textit{E}), west (\textit{W}), south (\textit{S}), and north (\textit{N}). 

\textit{Definition 2 (Traffic Lanes):} Traffic road networks typically comprise three distinct types of lanes: left-turn \textit{lef}, straight-through \textit{str}, and right-turn \textit{rig}. The traffic lanes in an intersection \textit{$I_i$} can be described as the upstream lanes \{{$X^i_{y}$}\} and downstream lanes \{{$X'^i_{y'}$}\}, in which \textit{$X,X'\in \{E,W,S,N\} $} and \textit{$y,y'\in \{lef,str,rig\} $}. Specifically, the vehicles enter the intersection \textit{$I_i$} are denoted as the upstream lanes \{{$X^i_{y}$}\}, and the vehicles leave the intersection \textit{$I_i$} are denoted as the downstream lanes \{{$X'^i_{y'}$}\}. 

\textit{Definition 3 (Traffic Movement):} Traffic movement is defined as the flow of traffic crossing an intersection from one upstream lane to one downstream lane. A traffic movement, such as from lane \textit{$X^{i}_y$} to lane \textit{$X'^i_{y'}$}, is denoted as (\textit{$X^{i}_y,X'^i_{y'}$}). At an intersection where each road comprises three lanes, one downstream lane has three upstream lanes to generate traffic movements, thereby a total of nine traffic movements for one road that includes three downstream lanes.

\textit{Definition 4 (Traffic Queue Length):} The traffic queue length \textit{$x(X^{i}_y)$} represents the number of vehicles waiting in lane \textit{$X^{i}_y$}.

\textit{Definition 5 (Traffic running vehicle number):} The traffic running vehicle number \(r(X^{i}_y)\) represents the number of vehicles running in lane \(X^{i}_y\).

\textit{Definition 6 (Traffic Signal Phase):} Each traffic signal phase represents a set of allowed traffic movements. The sum of phases in intersection \textit{$I_i$} is denoted as \textit{$J_i$}, and notation \textit{j} is used to denote one of the phases. Fig. \ref{fig:2}(b) demonstrates the mostly used eight phases.

\textit{Definition 7 (Efficient pressure):} The efficient pressure (EP) is the difference between the average queue length on upstream lanes and the average queue length on downstream lanes.
\begin{equation}
p_e({X^{i}_y}, {X'^i_{y'}}) = \frac{1}{L} \sum_{i=k}^{L} x(l_k) - \frac{1}{M} \sum_{j=1}^{M} x(m_j),
\end{equation}
where \(l_k\in \{X^{i}_y\}, m_j\in \{X'^i_{y'}\}\), \textit{L} and \textit{M} represent the number of lanes of \(\{X^{i}_y\}\) and \(\{X'^i_{y'}\}\). The schema is shown in the left part of Fig. \ref{fig:2}(a).

\textit{Definition 8 (Adaptive pressure):} The attention-based adaptive pressure (AP) is the difference between the queue length at each upstream lane and the weighted queue length on downstream lanes.
\begin{equation}
\label{equation11}
p_e(X^{i}_y, {X'^i_{y'}}) = x(l_k) - \sum_{j=1}^{M} \omega^k_j x(m_j),
\end{equation}
where \(l_k\in \{X^{i}_y\}, m_j\in \{X'^i_{y'}\}\), \textit{M} represents the number of lanes of \(\{X'^i_{y'}\}\), and \(\omega^k_j\) represents the weight of vehicle flow of upstream lane \(l_k\) on downstream lane \(m_j\). The schema is shown in the right part of Fig. \ref{fig:2}(a).

\subsection{Markov property in RL-based ATSC}
\label{sec:subsecIIB}
Due to the shared Markov property, ATSC can be effectively characterized as a Markov process, where state transitions depend only on the current state and are independent of historical states. This property enables ATSC to be effectively formulated as a Markov Decision Process (MDP). In ATSC, RL provides a powerful framework for solving ATSC problems by optimizing control policies to maximize the expected cumulative reward, leveraging the structure of ATSC scenarios. Therefore, at time \textit{t} in intersection \textit{$I_i$}, RL is composed of six fundamental elements: the state space \(\mathcal{S}\) = \textit{$\{s^1_t,\cdots,s^n_t\}$}, observation space \textit{O} = \textit{$\{o^1_t,\cdots,o^n_t\}$}, action space \textit{A} = \textit{$\{a^1_t,\cdots,a^n_t\}$}, transition probability function \textit{$P(s^{i'}_t|s^i_t,a^i_t)$}, reward function \textit{$R(s^i_t,a^i_t)$} = \textit{$\{r^1_t,\cdots,r^n_t\}$}, and discount factor \( \gamma \). 

The goal of the MDP formulation in ATSC is to make global optima policies \( \pi = \{\pi^1 _t,...,\pi^n_t\}\) for each intersection \textit{$I_i$}. These policies aim to maximize the own expected cumulative reward of taking a specific action \textit{$a_t^i$} in a given state \textit{$s_t^i$} over all future time steps, i.e., the state value function:
\begin{equation}
V(s_t^i) = \mathbb{E}_{\pi^i_t} \left[ \sum_{k=0}^{\infty} \gamma^k r_{t+k+1}^i \mid s_t^i \right],
\end{equation}
where \textit{$\pi^i_t$}: \textit{$o^i_t \times a^i_t \rightarrow [0,1]$} maps the observation of intersection \textit{$I_i$} to the probability distribution of its action. The Q-value (action-value) of each intersection is defined as 
\begin{equation}
Q(s_t^i, a_t^i) = \mathbb{E}_{\pi^i_t} \left[ \sum_{k=0}^{\infty} \gamma^k r_{t+k+1}^i \mid s_t^i, a_t^i \right].
\end{equation}

Q-learning is an effective RL algorithm that aims to find the optimal action-selection policy by iteratively updating Q-values based on the Bellman equation \cite{r43}. The update rule is expressed as
\begin{equation}
\label{equation4}
Q(s_t^i, a_t^i) \leftarrow Q(s_t^i, a_t^i) + \alpha \eta_t^i,
\end{equation}
where \textit{$\alpha$} is the learning rate and
\begin{equation}
\label{equation5}
\eta_t^i = \left[ r_t^i + \gamma \max_{a_{t+1}^i} Q(s_{t+1}^i, a_{t+1}^i) - Q(s_t^i, a_t^i) \right].
\end{equation}

For each intersection \textit{$I_i$}, the RL agent calculates the expected Q-values for eight phases. This process derives optimal policies to enhance traffic efficiency.

\begin{figure*}[t]
    \centering
    \includegraphics[width=1.03\textwidth, trim={1.35cm 0 0 0}, clip]{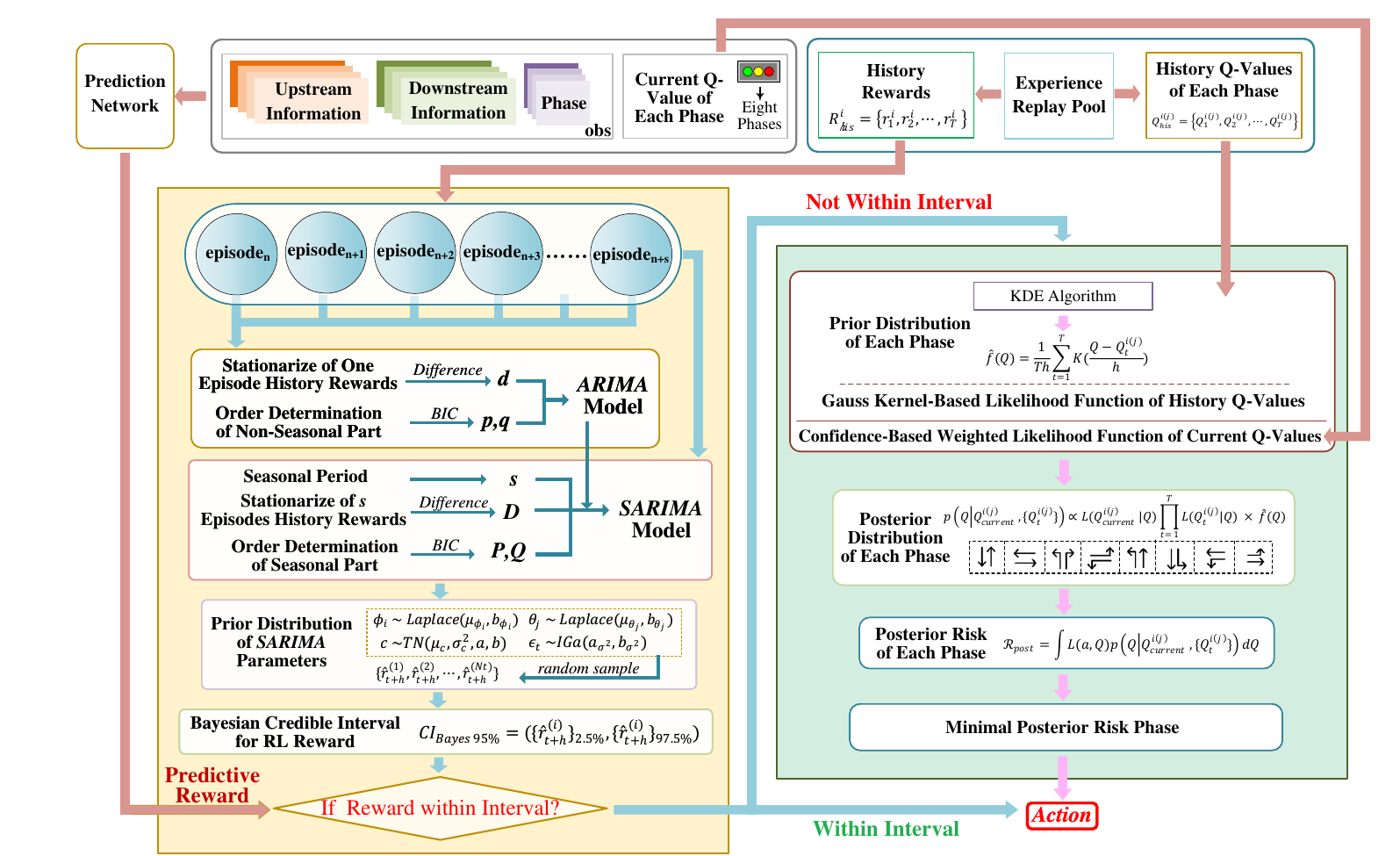}
    \caption{\raggedright Framework of Bayesian Critique-Tune for RL.}
    \label{fig:3}
\end{figure*}

\section{Bayesian Critique-Tune Framework for RL}
\label{sectionIII}

To achieve the refinement of excessive trust in RL policies, this paper designs a two-layer Bayesian structure to refine RL policies. Subsection \ref{sec:subsecIIIA} describes the overall description of structure and a prediction network. Subsection \ref{sec:subsecIIIB} provides an in-depth discussion of the Critique Layer and Subsection \ref{sec:subsecIIIC} presents the Tune Layer in detail.

\subsection{Prediction Network and Bayesian Critique-Tune Structure}
\label{sec:subsecIIIA}

\subsubsection{Prediction Network}
In the Bayesian Critique-Tune Framework framework, an auxiliary prediction network is implemented to predict the Predictive Reward \(\hat r^i_{t+h}\) corresponding to the next action. 

Specifically, the original environment states, including the vehicle queue length at each upstream and downstream lane, the number of running vehicles in each upstream and downstream lane, and the current phase of the intersection, are inputted to the prediction network. In addition, the current Q-value \(Q_{{cur}}^{i(j)}\), shared from the DQN, is also provided as an input. 

This paper employs a multilayer perceptron (MLP) to construct the prediction network. Equipped with three fully connected layers, the prediction network enables simple yet efficient supervised learning. Thereby capable of generating the suitable Predictive Reward.

\subsubsection{Bayesian Critique-Tune structure}
The Bayesian Critique-Tune structure includes two layers. The Bayesian inference-based Critique Layer constructs a credible interval \(CI_{\text{Bayes}}\) by utilizing historical rewards \textit{$R^i_{his}$} = \textit{$\{r^i_1,r^i_2,\cdots,r^i_T\}$} and Bayesian prior experience. This interval evaluates whether the Predictive Reward \(\hat r^i_{t+h}\) falls within an acceptable range based on prior knowledge. Subsequently, the Bayesian decision-based Tune Layer is activated if the evaluation yields a negative outcome, which meaning that \( \hat r^i_{t+h}\notin CI_{\text{\textit{Bayes}}} \). For each phase, the Tune Layer computes the posterior risk \(\mathcal{R}_{\text{\textit{post}}}\) by integrating the posterior distribution \( p(Q | Q_{{his}}^{i(j)}, Q_{{cur}}^{i(j)}) \) of the history Q-values \(Q_{{his}}^{i(j)}\) = \textit{$\{Q^{i(j)}_1,Q^{i(j)}_2,\cdots,Q^{i(j)}_T\}$}  and current Q-value \(Q_{{cur}}^{i(j)}\), in which \(j = 1,2,\cdots,8 \) denoted as each phase. Via selecting the phase with the minimal posterior risk \( \min \mathcal{R}_{\text{\textit{post}}}\), the Tune Layer ensures the updated policy \( \pi_{upd}^i \)aligned with the global optima. Thereby avoid the problem of  excessive trust of RL policies.

\subsection{Critique Layer}
\label{sec:subsecIIIB}

The Bayesian inference-based Critique Layer is responsible for evaluating the reasonableness of RL policies. Its judging process is divided into three main sections.

\subsubsection{SARIMA Modeling of History Rewards} The history rewards \textit{$R^i_{his}$} of RL are subjected to rigorous time series analysis. Thus, it can be mathematically represented using a time series model. Initially, the Augmented Dickey-Fuller (ADF) test is employed to judge the stationarity of the history rewards data. If the data exhibits non-stationarity, an appropriate \textit{differencing} process is applied. Typically, the expression for a \textit{d-th} order \textit{differencing} is given by:
\begin{equation}
\label{equation7}
w_t^{i(d)} = (1 - B)^d r_t^i,
\end{equation}
where \textit{$r_t^i$} and \textit{$w_t^{i(d)}$} represent the original and \textit{differenced} history rewards data, \textit{B} is the lag operator that defined as \textit{$B r_t^i = r_{t-1}^i$}, and \textit{d} denotes the \textit{differencing} order.

Due to the long-term dependence properties within a single episode, RL-based ATSC can be modeled using an Autoregressive Integrated Moving Average (ARIMA) model. Each episode’s data can then be represented by the ARIMA equation:
\begin{equation}
\label{equation3}
(1 - \sum_{k=1}^{p} \phi_k B^k)(1 - B)^d r_t^i = (1 + \sum_{j=1}^{q} \theta_j B^j) \epsilon_t,
\end{equation}
where \textit{$\phi_i$} represents the autoregressive coefficients, \textit{$\theta_j$} represents the moving average coefficients, \textit{$\epsilon_t$} represents the random error term at time \textit{t} and typically assumed to be white noise (i.e. \textit{$\epsilon_t \sim \mathcal{N}(0, \sigma^2)$}). Moreover, \textit{p} denotes the order of the autoregressive (AR) part, and \textit{q} denotes the order of the moving average (MA) part.

Subsequently, the optimal orders \textit{$\{p,d,q\}$} of the ARIMA model need to be determined, tailored to the complexities inherent in RL-based ATSC. The determination process uses the Bayesian Information Criterion (BIC) to evaluate and select the model order, balancing complexity and predictive accuracy. The BIC formulation is given as follows:
\begin{equation}
\label{equation9}
\text{BIC} = -2 \ln(L) + k \ln(n),
\end{equation}
where \(L\) is the likelihood function value of the model, and \(k\) represents the number of parameters in the ARIMA model.

In order to enhance the rationality of the credible interval, this paper introduces the Seasonal Autoregressive Integrated Moving Average (SARIMA) model to fit the multi-episodes data, with each episode regarded as the same ARIMA model (the detailed theoretical explanation is presented in Appendix \ref{proof 1}). In detail, the SARIMA model combines non-seasonal and seasonal components, which are the non-seasonal part ARIMA\((p, d, q)\) and the seasonal part SARIMA\((P, D, Q, s)\). The non-seasonal part of ARIMA (p, d, q) is defined as above. The definitions of the seasonal part of SARIMA\((P, D, Q, s)\) are as follows: \textit{P} denotes the order of the seasonal autoregressive part, \textit{Q} denotes the order of the seasonal moving average part, \textit{D} denotes the seasonal \textit{differencing} order, and \textit{s} denotes the seasonal period. The optimal orders \textit{$\{P,D,Q\}$} also determined by Eq. (\ref{equation9}).

Building on the above methods, the SARIMA equation for historical rewards in the traffic signal control context can be modeled:
\begin{equation}
\label{equation8}
\Phi_P(B^s) \phi_p(B) \nabla^d \nabla_s^D r_t^i = \Theta_Q(B^s) \theta_q(B) \epsilon_t,
\end{equation}
where \(\phi_p(B) = 1 - \phi_1 B - \phi_2 B^2 - \cdots - \phi_p B^p
\), \(\Phi_P(B^s) = 1 - \Phi_1 B^s - \Phi_2 B^{2s} - \cdots - \Phi_P B^{Ps}
\), \(\nabla^d r_t^i = (1 - B)^d r_t^i\), \(\nabla_s^D r_t^i = (1 - B^s)^D r_t^i\), \(\theta_q(B) = 1 + \theta_1 B + \theta_2 B^2 + \cdots + \theta_q B^q\), \(
\Theta_Q(B^s) = 1 + \Theta_1 B^s + \Theta_2 B^{2s} + \cdots + \Theta_Q B^{Qs}
\).

Establishing this SARIMA model enables an accurate quantity of RL policies by capturing temporal patterns of the data. Achieving the effectiveness of the follow-up evaluation process of the Predictive Reward.

\subsubsection{Bayesian Prior Distribution of SARIMA Parameters} By integrating the prior experience of traffic data properties, the SARIMA model achieves effective evaluation for the RL policies in the next step. Firstly, the determination of the prior distribution of unknown parameters \(\boldsymbol{\xi} = (c, \phi_1, \ldots, \phi_p, \theta_1, \ldots, \theta_q, \sigma^2)\) is a pivotal step for this section. Specifically, due to the non-negativity of history rewards \textit{$R^i_{his}$}, the Truncated Normal distribution \textit{$\text{TN}(\mu_c, \sigma_c^2, a, b) $} is suitable for the intercept term \textit{c}. Where \textit{$\mu_c$} is the mean of the untruncated normal distribution, \textit{$\sigma^2_c$} is the variance of the untruncated normal distribution, \textit{a} and \textit{b} are the lower and upper bounds of the truncation interval, respectively. And due to the complexity of the traffic environment, incorporating a sparsity-inducing prior is essential when modeling history rewards \textit{$R^i_{his}$}. Therefore, the Laplace distribution \textit{$ Laplace(\mu_{\phi_i}, b_{\phi_i}) $} and Laplace distribution \textit{$ Laplace(\mu_{\theta_j}, b_{\theta_j}) $} are selected for autoregressive coefficients \textit{$\phi_i$} and moving average coefficients \textit{$\theta_j$}, respectively. Where \(\mu_{\phi_i}\) and \(\mu_{\theta_j}\) are the means, \(b_{\phi_i}\) and \(b_{\theta_j}\) are the scale parameters, with a smaller \textit{b} indicating a stronger tendency toward coefficient sparsity. Furthermore, the random error term \textit{$\epsilon_t$} in RL has large uncertainties, thereby the Inverse Gamma distribution \textit{$IGa(a_{\sigma^2},b_{\sigma^2})$} is applied to estimate the variance \textit{$\sigma^2$} of the random error term \textit{$\epsilon_t$}. Where \(a_{\sigma^2}\) is the shape parameter and \(b_{\sigma^2}\) is the scale parameter. The heavy-tailed nature of this distribution effectively captures the potential for larger variance values. Above all, the prior distribution can be expressed as follows:
\begin{equation}
p(c) = \frac{1}{\sigma_c} \cdot \frac{\phi\left(\frac{c - \mu_c}{\sigma_c}\right)}{\Phi\left(\frac{b - \mu_c}{\sigma_c}\right) - \Phi\left(\frac{a - \mu_c}{\sigma_c}\right)},
\end{equation}
\begin{equation}
p(\phi_i) = \frac{1}{2b_{\phi_i}} \exp\left(-\frac{|\phi_i-\mu_{\phi_i}|}{b_{\phi_i}}\right),
\end{equation}
\begin{equation}
p(\theta_j) = \frac{1}{2b_{\theta_j}} \exp\left(-\frac{|\theta_j-\mu_{\theta_j}|}{b_{\theta_j}}\right),
\end{equation}
\begin{equation}
p(\sigma^2) = \frac{b_{\sigma^2}^{a_{\sigma^2}}}{\Gamma(a_{\sigma^2})} (\sigma^2)^{-a_{\sigma^2}-1} \exp\left(-\frac{b_{\sigma^2}}{\sigma^2}\right),
\end{equation}
where \(\boldsymbol{\xi} = (c, \phi_1, \ldots, \phi_p, \theta_1, \ldots, \theta_q, \sigma^2)\) are all the parameters to be estimated, and \(\Phi(x) = \frac{1}{\sqrt{2\pi}} e^{-\frac{x^2}{2}}\)

After obtaining the prior distribution of each parameter, the sample set \( \{\boldsymbol{\xi}^{(1)}, \boldsymbol{\xi}^{(2)}, \ldots, \boldsymbol{\xi}^{(Nt)}\} \) can be generated by random sample. And the values set \(\{\hat{r}_{t+h}^{(i)}\}\) using each sample \( \{\boldsymbol{\xi}^{(i)} \) at future time \(t+h\) can be calculated by Eq. (\ref{equation8}).

The above process achieves effective estimation of the SARIMA parameters. Despite the high levels of uncertainty and instability present in history rewards data \textit{$R^i_{his}$}, this process enables the acquisition of reliable sampled RL reward values set\(\{\hat{r}_{t+h}^{(i)}\}\).In complex traffic environments, the history rewards always fail to serve as robust statistical metrics. Thereby the reasonableness of the RL policies is struggled to evaluate by traditional methods. Conversely, by leveraging the prior experience, the influence of these limitations can be effectively addressed. The above process establishes a foundation for conducting a reasonable and comprehensive evaluation of RL policies in subsequent analyses. Thereby enhancing the reliability and effectiveness of RL policies for ATSC.

\subsubsection{Bayesian Credible Interval for RL Reward} The Bayesian credible interval for RL reward values \textit{$\hat{r}_{t+h}^{(i)}$} is constructed to evaluate the reasonableness of the RL policies. Based on the RL reward values set\(\{\hat{r}_{t+h}^{(1)}, \hat{r}_{t+h}^{(2)}, \ldots, \hat{r}_{t+h}^{(Nt)}\}\), the Bayesian credible interval is calculated by the percentiles of this sample set. Specifically, this paper constructs a \(95\%\) Bayesian credible interval; take the \(2.5th\) and \(97.5th\) percentiles of the posterior sample values set:
\begin{equation}
\text{\textit{CI}}_{Bayes95\%} = \left(\{\hat{r}_{t+h}^{(i)}\}_{2.5\%}, \{\hat{r}_{t+h}^{(i)}\}_{97.5\%} \right),
\end{equation}
This interval represents the range within which the future value \(r_{t+h}^{(i)}\) falls with a \(95\%\) probability, given the observed data and the prior information.

Building on the above interval, the Predictive Reward \(\hat r^i_{t+h}\) is generated by the prediction network can be evaluated through the Bayesian credible interval. If the Predictive Reward falls within the range of the Bayesian credible interval, the RL policies will be employed. Conversely, the policies need to be fine-tuned.

\subsection{Tune Layer}
\label{sec:subsecIIIC}

The Bayesian decision-based Tune Layer is responsible for fine-tuning RL policies by posterior risk when the evaluation is negative. This layer ensures the RL policies adapt to real-time traffic conditions, enhancing the reasonableness of the policy-making process. The tuning process is divided into two main sections.

\subsubsection{The Incorporating of Q-values Information} This part main involves determining the prior distribution of history Q-values \textit{$Q^{i(j)}_{his}$} and constructing the Q-values \(\{Q^{i(j)}_{cur}, Q^{i(j)}_{his}\}\) likelihood function. Firstly, given the high entropy inherent in traffic environments, Q-values often exhibit multiple peaks and may present a noninformative prior. For this complexity, this paper employs the Kernel Density Estimation (KDE), a nonparametric Bayesian method, to estimate the probability density function of history Q-values. This estimated density is then utilized to construct the prior distribution:
\begin{equation}
\hat{f}(Q) = \frac{1}{Th} \sum_{t=1}^{T} K\left(\frac{Q - Q^{i(j)}_t}{bw}\right),
\end{equation}
where \( bw \) is the bandwidth parameter, \( K(u) \) is the Gauss kernel, in which \(K(u) = \frac{1}{\sqrt{2\pi}} exp({-\frac{u^2}{2}})\)

Concurrently, the likelihood function of Q-values can be constructed, which includes two parts. Specifically, based on the Gauss kernel, the likelihood function of the history Q-values \textit{$Q^{i(j)}_{his}$} is defined as the follows:
\begin{align}
    L(Q^{i(j)}_t | Q) &= K\left(\frac{Q - Q^{i(j)}_t}{bw}\right)\nonumber \\
                & = \frac{1}{\sqrt{2 \pi {bw}^2}} \exp\left(-\frac{(Q - Q^{i(j)}_t)^2}{2{bw}^2}\right),
\end{align}
Additionally, since the current Q-value \(Q^{i(j)}_{cur}\) carries greater significance in RL-based policy-making process, a confidence-based weighting mechanism is designed. This paper uses the normal distribution with an error term to describe the likelihood of the current Q-value, denoted as the "weighted likelihood function":
\begin{equation}
L(Q_{{cur}}^{i(j)} | Q) = \frac{1}{\sqrt{2\pi\sigma^2_{cur}}} \exp\left(-\frac{(Q - Q_{{cur}}^{i(j)})^2}{2\sigma^2_{cur}}\right),
\end{equation}
where \( \sigma^2_{cur} \) is a tuning parameter representing the uncertainty of the current Q-value. A smaller \( \sigma^2_{cur} \) value means higher confidence in the current Q-value. This adjustment allows the Tune Layer to prioritize the most recent state information, which is crucial for capturing the dynamic nature of traffic flow.\\
By combining the weighted likelihood of the current Q-value \( L(Q_{{cur}}^{i(j)} | Q) \) with the history Q-value prior distribution \( \hat{f}(Q)\), the overall likelihood function is expressed as follows:
\begin{equation}
L(Q) = L(Q_{{cur}}^{i(j)} | Q) \times \prod_{t=1}^{T} L(Q^{i(j)}_t|Q),
\end{equation}
This integrated likelihood function enhances the adaptability and responsiveness of the RL agent to real-time traffic conditions. Ensuring comprehensive and effective subsequent Bayesian-based phase tuning.

\subsubsection{Bayesian Posterior Risk of Each Phase} As in the Critique Layer, the Bayesian posterior distribution is initially updated according to \textit{Bayes’ theorem}:
\begin{equation}
p(Q | Q_{{cur}}^{i(j)}, \{Q^{i(j)}_t\}) \propto L(Q) \times \hat{f}(Q).
\end{equation}

Subsequently, based on the Bayesian decision theory, a loss function \( L(Q_{{cur}}^{i(j)}, Q) \) is introduced to measure the posterior risk(this paper adopts the square loss function \(L(Q_{{cur}}^{i(j)}, Q) = (Q_{{cur}}^{i(j)} - Q)^2
\)). The expected posterior risk for each phase in the intersection \textit{$I_i$} is expressed as follows:
\begin{equation}
\mathcal{R}_{\text{\textit{post}}} = \int_{\Theta}  L(Q_{{cur}}^{i(j)}, Q) \, p(Q | Q_{{cur}}^{i(j)}, \{Q^{i(j)}_t\}) \, dQ.
\end{equation}

Building on the Bayesian criteria,  the posterior risk for each phase is computed, and the phase that minimizes the expected posterior risk \(\mathcal{R}_{\text{\textit{post}}} \) is selected as the optimal action:
\begin{equation}
  min\mathcal{R}_{\text{\textit{post}}} = \inf_j \int_{\Theta}  L(Q_{{cur}}^{i(j)}, Q) \, p(Q | Q_{{cur}}^{i(j)}, \{Q^{i(j)}_t\}) \, dQ.
\end{equation}

Under the square loss function \(L(Q_{{cur}}^{i(j)}, Q) = (Q_{{cur}}^{i(j)} - Q)^2\), it can be theoretically proven that the above process minimizes the posterior risk (the detailed proof is presented in Appendix \ref{proof 2}). Thereby leading to the most effective policy refinement for RL. This methodology ensures the selected phase achieves the lowest total policy-making risk, enhancing the reasonableness of the RL policies for ATSC.

\begin{figure*}[t]
    \centering
    \includegraphics[width=\textwidth]{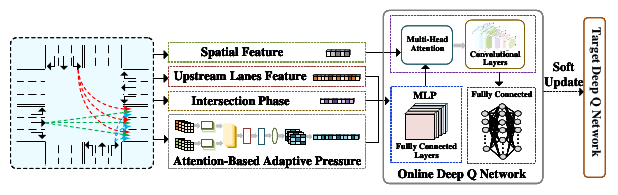}
    \caption{\raggedright Architecture of attention-based adaptive pressure extraction for each intersection direction.}
    \label{fig:4}
\end{figure*}

\section{RL With Attention-Based Adaptive Pressure for ATSC}
\label{sectionIV}

This section presents the attention-based adaptive pressure RL for ATSC. In Subsection \ref{sec:subsecIVA}, an Attention-Based Adaptive Pressure (AP) is introduced. Subsection \ref{sec:subsecIVB} details the AP-based DQN for ATSC, which serves as the policy-making backbone of the total framework. Finally, Subsection \ref{sec:subsecIVC} outlines the policy-making and training process for the proposed BCT-APLight.

\subsection{Attention-Based Adaptive Pressure Extraction}
\label{sec:subsecIVA}

In the complex urban environment, effective and efficient traffic movement representation of each lane is crucial for optimizing signal control. To achieve effectively capture the real-time traffic features, this paper proposes an Attention-Based Adaptive Pressure. 

In an intersection \textit{$I_i$}, the AP mechanism identifies the source upstream lane associated with each vehicle and links it to the corresponding downstream lane. The downstream lanes and upstream lanes are represented as four-layer matrices of dimensions \(3\times 3\), where the matrices are denoted as \textit{K} and \textit{Q}, respectively. Each of the three columns in these matrices represents: the number of waiting vehicles \(k_{wai}\) and \(q_{wai}\), the number of running vehicles \(k_{run}\) and \(q_{run}\), and the total number of vehicles \(k_{tot}\) and \(q_{tot}\). The three rows correspond to the left-turn lane \textit{lef}, straight-through lane \textit{str}, and right-turn lane \textit{rig}. Besides, the four layers of the matrix represent the directions: east \textit{E}, west \textit{W}, south \textit{S}, and north \textit{N}.

The AP mechanism further utilizes a multi-head attention mechanism to process and extract detailed information from both downstream and upstream lanes. This processed data are subsequently used as inputs for the query matrix \(Q\) and key matrix \(K\). The multi-head attention mechanism is formalized as follows:
\begin{equation}
W_u^i = \psi_{\text{linear}} (\text{Concat}(W_1^i, W_2^i, \ldots, W_{head}^i)).
\end{equation}
where,
\begin{equation}
W_l^i = \text{Softmax}\left(\frac{q_l^i (K_l^i)^T}{\sqrt{d_k^i}}\right).
\end{equation}
In this formulation, \(head\) represents the number of attention heads, and \(d_k\) is the dimensionality of the key \(K\). This multi-head attention mechanism allows the AP system to construct a four-layer upstream attention weight matrix with dimensions \(3\times 3\). This matrix represents the weight of vehicle number influence of each upstream lane on three downstream lanes. A higher attention weight signifies a stronger impact. Based on these attention weights and Eq. (\ref{equation11}), the AP mechanism can be obtained. Unlike traditional methods that treat the queue lengths of all downstream lanes as coming equally from each upstream lane. AP achieves adaptability and effective capture of the traffic features, avoiding the lackness of traditional methods. Wherein these weaknesses may amplify the impact of low-traffic lanes while weakening responsiveness to high-traffic ones.

\subsection{AP-based DQN for ATSC}
\label{sec:subsecIVB}

In the ATSC, this paper adapts the RL method using Deep Q-Networks (DQN), tailored to optimize the traffic signal phase policy-making process at intersections. The AP-based DQN algorithm employs artificial neural networks (ANNs) to approximate the optimal action-value function \(Q(s_t^i, a_t^i)\) for each phase selection at each time step \(t\). The architecture of AP-based DQN for ATSC is shown in Fig. \ref{fig:4}. This dual-network framework comprises both an online network and a target network, with parameters \( \zeta \) and \( \zeta^{*}\) respectively. These two networks maintain the same structure but are updated at different rates to stabilize the learning process. 

For AP-based DQN signal control in an intersection \(I_i\), the state \(s^i_t\) is indicated as the current traffic conditions, which include the AP, the intersection phase, the number of running vehicles of upstream lanes, and the spatial correlation of each intersection. These state information forms the basis upon which the DQN estimates \(Q^{\pi_t^i} \left(s^i_t | \zeta \right)\), enabling the system to select the optimal traffic signal phase \(j \in J_i\) for minimizing congestion and delay at the intersection. Besides, the action space \textit{A} consists of predefined signal phases, and the objective is to maximize the cumulative reward by optimizing the traffic flow. The reward \(r^i_t\) is designed to reflect traffic efficiency, integrating both a phase-switching penalty and a traffic throughput incentive. A negative reward is imposed for unnecessary phase switching. In contrast, a positive reward is provided for improvements in transportation efficiency.

By interacting with experience replay pool \(\omega\), DQN improves learning efficiency by storing all replay experience \(\{\{(s^i_t, a^i_t, r^i_t, s^i_{t+1}),\cdots,(s^i_{st}, a^i_{st}, r^i_{st}, s^i_{st+1})\}\}_{i=1}^n\) of one epoch \textit{st}, and computing the following gradient by differentiating the loss function with respect to the weights:
\begin{equation}
\label{equation6}
\nabla_{\zeta} L(\zeta) = \alpha \eta_t \nabla_{\zeta} Q(s^{*}_t, a^{*}_t; \zeta),
\end{equation}
where \(\alpha\) denotes the learning rate, \(s^{*}_t, a^{*}_t\) belongs to the target network, and the temporal-difference (TD) error \(\eta_t = \{\eta_t^{i^{*}},\cdots,\eta_t^{n^{*}}\}\) is calculated as:
\begin{equation}
\eta_t^{i^{*}} = r^{i^{*}}_t + \gamma \max_{a^{i^{*}}_{t+1}} Q(s^{i^{*}}_{t+1}, a^{i^{*}}_{t+1}; \zeta^{*}) - Q (s^{i^{*}}_t, a^{i^{*}}_t; \zeta),
\end{equation}
The online network parameters\(\zeta\) are updated by stochastic gradient descent and Eq. (\ref{equation6}). And the target network parameters \(\zeta^{*}\) are updated via soft updates:
\begin{equation}
\zeta^{*} \leftarrow \tau \zeta + (1 - \tau) \zeta^{*}.
\end{equation}

In this algorithm, the AP-based DQN framework effectively adapts to ATSC. Equipped with the AP mechanism, the system achieves a more refined and context-aware signal control. This adaptive algorithm minimizes average travel time at urban intersections, making the RL policies efficiently responsive to fluctuating traffic conditions. 

\renewcommand{\thealgocf}{1} 
\begin{algorithm}
    \SetAlgoLined 
	\caption{BCT-APLight Policy-Making and Training Algorithm for Multi-Intersections}
    \label{alg:BCT-APLight}
	\KwIn{DQN-based agents for multi-intersections \( \{I_i\}_{i=1}^n \) ATSC;}
	\KwOut{Optimized signal phase policies for all intersections \( \{ \pi_{\textit{upd}}^i \}_{i=1}^n \);}
    Initialize experience replay pool \(\omega\)\;
	Initialize online network \( \zeta \), target network \( \zeta^{*} \leftarrow \zeta \), and auxiliary prediction network\;
	\For{episode $= 1$ to max\_epoch}{
		\For{t $= 1$ to $T$}{
			Initialize traffic states $s_0^i$ for intersection $I_i$\;
            \If{\(episode > 10 \)}{
					Obtain \(R^i_{his}\) and \(Q^{i(j)}_{his}\) from \(\epsilon^i \)\;
                    \For{\( I_i \) in \( \{I_i\}_{i=1}^n \)}{
				Obtain \( a_t^i \) by \(Q^{i(j)}_{cur}\);
				Obtain \( \hat r^i_{t+h} \)\;
				
				\textbf{Critique Layer:}  Construct \( CI_{\text{Bayes}} \);
			    Evaluate if \( \textit{$\hat r^i_{t+h}$} \in CI_{\text{Bayes}} \)\;
			    \If{\( \textit{$\hat r^i_{t+h}$} \notin CI_{\text{Bayes}} \)}{
					\textbf{Tune Layer:} Update \( \pi_{\textit{upd}}^i \) by obtaining the phase with \( \min \mathcal{R}_{\text{post}} \)\;
				}
                }
			}
            Observe \( s_{t+1}^i \) and \( r_t^i \)\;
            Store \( (s_t^i, a_t^i, r_t^i, s_{t+1}^i) \) in \(\omega\)\;
            Update \( \zeta \) via \(\nabla_{\zeta} L(\zeta)\)\;
		}
	\If{Done}{
		\textbf{break}\;
				}
	}
	\sloppy
    \Return Optimal \( \{ \pi_{\textit{upd}}^i \}_{i=1}^n \) of intersections;
    \unskip
\end{algorithm}

\subsection{BCT-APLight Policy-Making and Training Process}
\label{sec:subsecIVC}
The policy-making and training process of the BCT-APLight for multi-intersections is illustrated in Algorithm \ref{alg:BCT-APLight}. Firstly, a DQN-based agent for ATSC is employed for multiple intersections \(\{I_i\}_{i=1}^n\). With the initialization of the experience replay pool \(\omega\), the online network \(\zeta\), target network \(\zeta^* \leftarrow \zeta\), and auxiliary prediction network are initialized for all intersections \(\{I_i\}_{i=1}^n\). For each intersection, the algorithm follows an episodic method to facilitate learning, updating the networks iteratively.

During each episode, the algorithm initializes the traffic states \(s_0^i\), retrieves history rewards \(R^i_{his}\) and history Q-values \(Q^{i(j)}{his}\) from the experience replay pool \(\omega\). Subsequently, for each time step \textit{t} within each episode, the current Q-values \(Q^{i(j)}{cur}\) are obtained using the \(\epsilon\)-greedy policy to select the most suitable action \(a_t^i\). Concurrently, the Predictive Reward \(\hat r^i_{t+h}\) is predicted using the auxiliary prediction network, providing a forecast for next performance.

Within each time step, the Critique Layer is employed to evaluate the credibility of the predicted reward. The history rewards \(R^i_{his}\) are used to construct a Bayesian credible interval \(CI_{\text{Bayes}} \). If the predicted reward \(\hat r^i_{t+h}\) falls outside this credible interval, the algorithm activates the Tune Layer. This layer updates the policy by calculating posterior risk \(\mathcal{R}{\text{post}}\) using the current Q-values \(Q{{cur}}^{i(j)}\) and the history Q-values \(Q_{{his}}^{i(j)}\), selecting the signal phase with the lowest risk.

The agent then observes the next state \(s_{t+1}^i\) and corresponding reward \(r_t^i\), which is subsequently used to update the online network via gradient descent. Besides, the agent stores the transition tuple \((s_t^i, a_t^i, r_t^i, s_{t+1}^i)\) in each replay buffer, ensuring the necessary experience is recorded for future use. After processing all time steps, the target network \(\zeta^*\) is updated based on the online network, ensuring stability in the learning process.

At the end of the process, the global optimal signal phase RL policies \(\{ \pi_{\textit{upd}}^i \}_{i=1}^n\) for all intersections are returned, culminating the training phase.

\section{Experiments and Results}
\label{sectionV}

To empirically evaluate the BCT-APLight, extensive experiments are conducted. Subsection \ref{sec:subsecVA} describes the overall experiment settings. Subsection \ref{sec:subsecVB} provides the analysis of experiment results in detail. The ablation study is shown in the \ref{sec:subsecVC}.

\subsection{Experiment Setup}
\label{sec:subsecVA}
\subsubsection{Environment Settings} This paper conducts experiments on the CityFlow traffic simulator \cite{r45}, which is open-sourced and simulates the kind of data that is collected at real-world intersections for ATSC. In the environment, each road has three lanes, which the number 0 corresponds to the left-turn lane, number 1 corresponds to the straight-through lane, and number 2 corresponds to the right-turn lane. Following most existing methods, the minimum action duration is set at 30 seconds, and a three-second yellow signal and a two-second all-red time follow each green signal to prepare the transition. Besides, this paper employs an epsilon decay strategy modeled as a power function, with a minimum epsilon value set at 0.2 during the training phase. Following the completion of training, epsilon decay is disabled (epsilon decay set to 0) for model evaluation in the testing phase.

\begin{figure}[ht]
    \centering
    \subfloat[Jinan]{\includegraphics[width=0.14\textwidth,height=2.5cm]{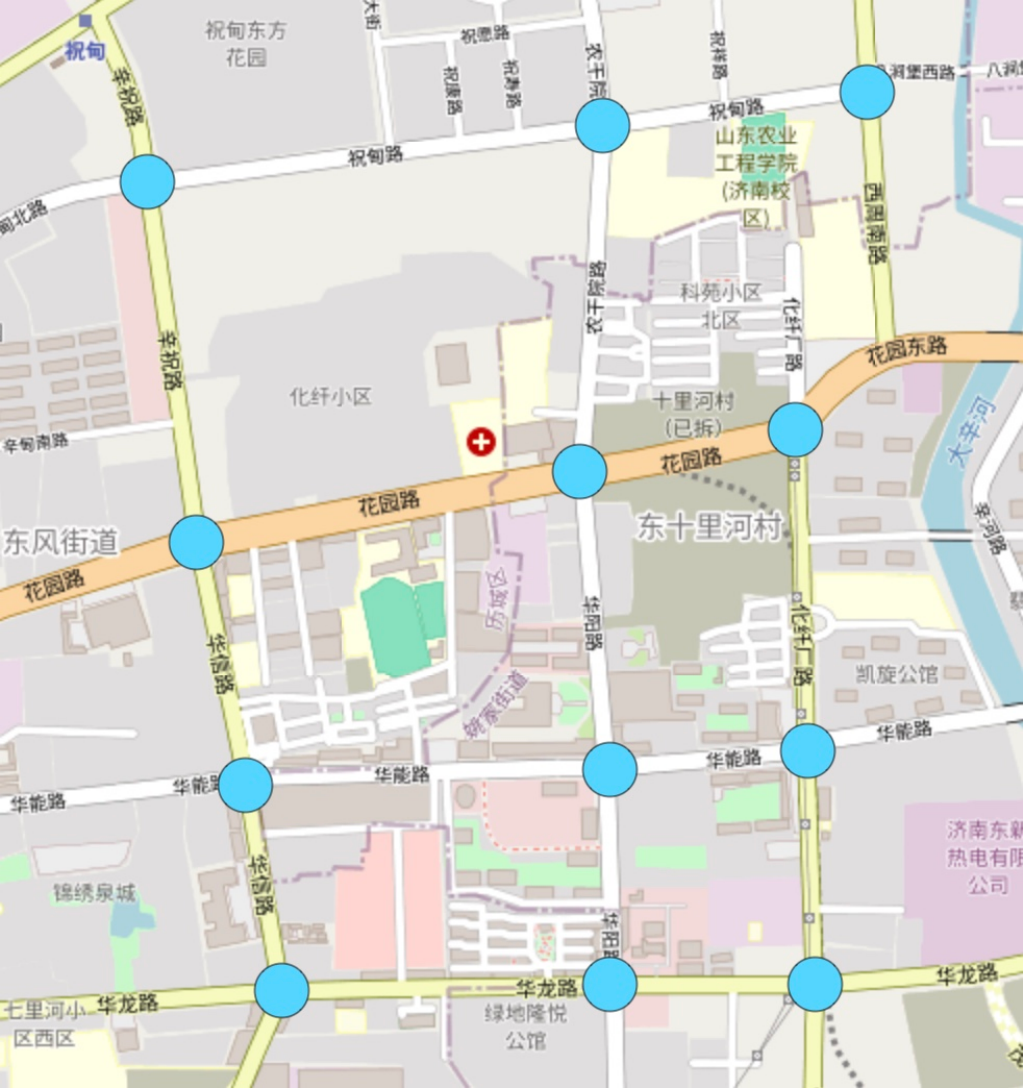}}
    \hspace{0.3cm}
    \subfloat[Hangzhou]{\includegraphics[width=0.14\textwidth,height=2.5cm]{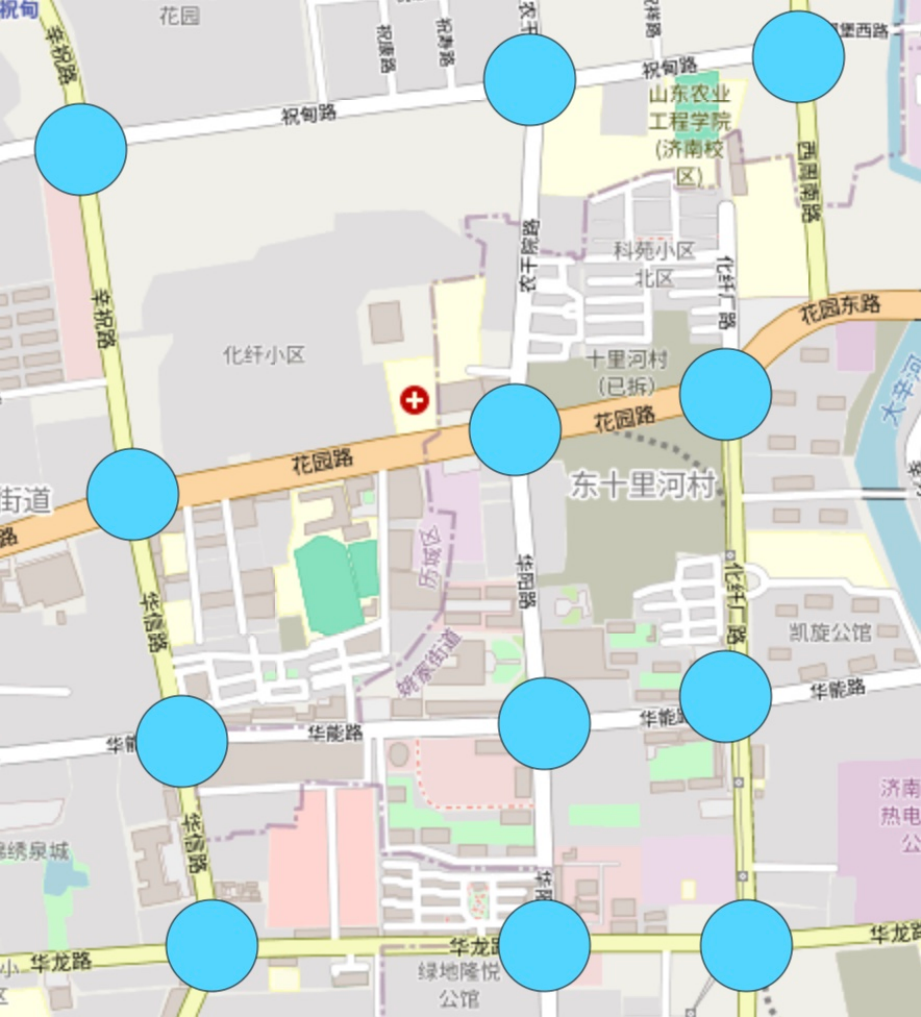}}
    \hspace{0.3cm}
    \subfloat[New York]{\includegraphics[width=0.14\textwidth,height=2.5cm]{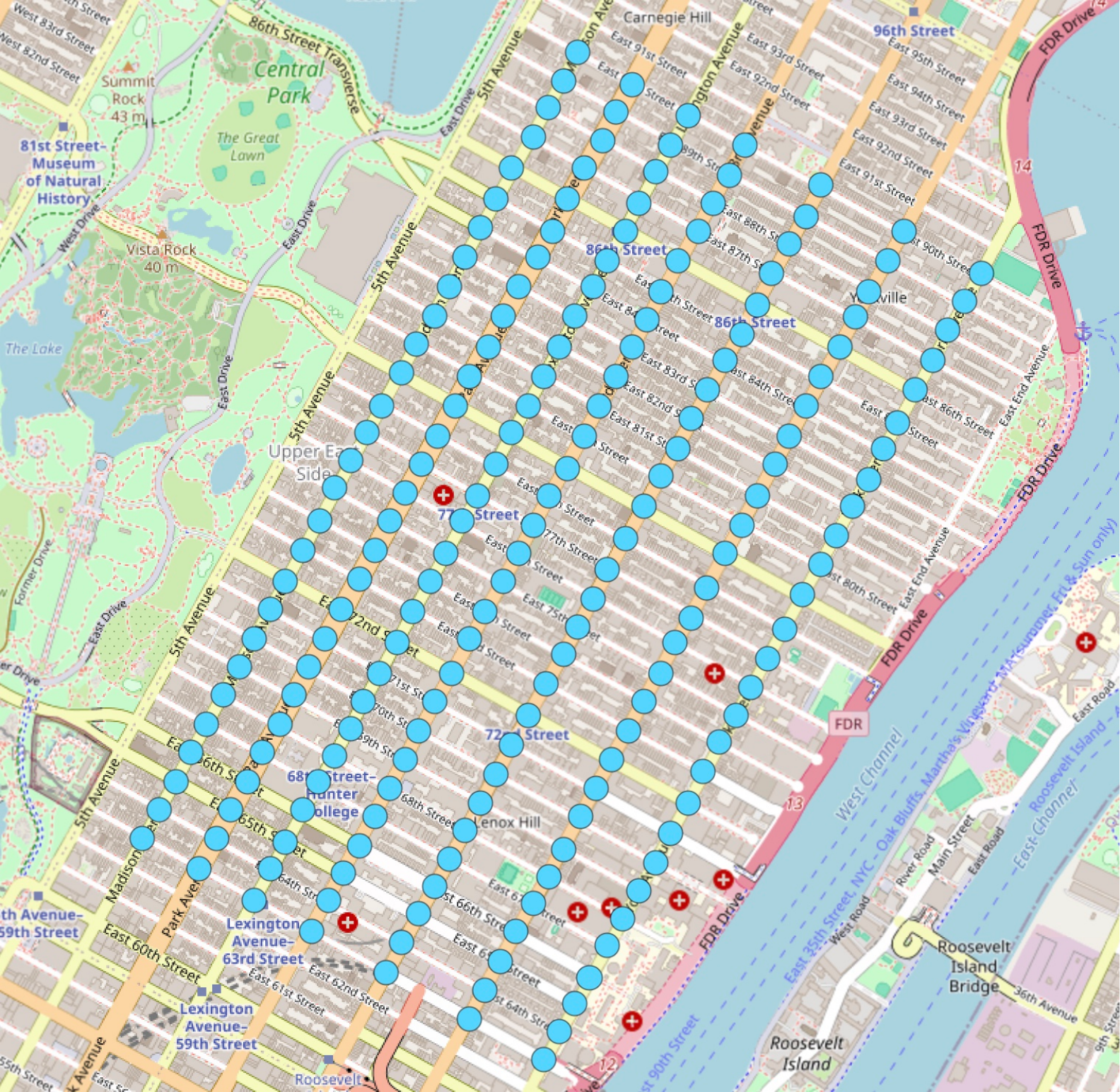}}
    \caption{The road network systems of the datasets from Jinan, Hangzhou, and New York, with uniform dimensions. The blue dots mark the traffic signal lights controlled by RL-agent.}
    \label{fig:5}
\end{figure}

\subsubsection{Datasets} The experiments are conducted on seven real-world traffic flow datasets, including three datasets in Jinan(\(3\times4\)), two datasets in Hangzhou(\(4\times4\)), and two datasets in New York(\(28\times7\)). There are four directions (E,W,S,N) at each intersection. Each direction has an incoming road and an outgoing road, and each road has three lanes: left-turn lane, straight-through lane, right-turn lane. The visualization of three areas is shown in Fig. \ref{fig:5}.
\subsubsection{Baselines} For evaluating the effectiveness of BCT-APLight, this paper compares it with three transportation methods, six RL-based methods, and one LLM-based method.

Transportation methods:
\begin{itemize}
    \item Random: A baseline method switches signal phases at random with a fixed duration.
    \item FixedTime\cite{r46}: Generally used in the majority of traffic situations, a method provides a pre-defined set cycle duration with phase time.    
    \item MaxPressure\cite{r47}: The SOTA ATSC method in the traditional transportation filed, maximizing traffic throughput by giving preference to the signal phase with the most traffic pressure.
\end{itemize}
RL-based methods:
\begin{itemize}
    \item MPLight\cite{r48}: A RL-based method that utilizes pressure as both observation and reward, and extends FRAP method as its foundational model.
    \item AttendLight\cite{r49}: Introducing attention mechanism to state space and action space of RL.
    \item PressLight\cite{r50}: Optimizing the pressure of the intersection by utilizing the MaxPressure concept with DRL.
    \item CoLight\cite{r51}: Employing the graph attention network among intersections, to capture the neighborhood information and enhance the coordination ability of the RL agent.
    \item Efficient-CoLight\cite{r52}: Building upon the CoLight model. Introducing the concept of efficient pressure, which is noticing the equal length of entering lanes.
    \item Advanced-CoLight\cite{r53}: Building upon the CoLight model. Considering both the running and waiting vehicles to incorporate advanced traffic states features.
\end{itemize}
LLM-based method:
\begin{itemize}
    \item LightGPT\cite{r54}: Introducing the Large Language Models into ATSC, leveraging the reasoning and decision-making process akin to human intuition for effective traffic control.
\end{itemize}
\subsubsection{Metrics} Following previous studies\cite{r53}, this paper leverages average travel time (ATT), average waiting time (AWT), and average queue length (AQL) of vehicles to evaluate the performance of each method.

\begin{itemize}
    \item Average traveling time (ATT): The average traveling time measures the average duration of all the vehicles traveling from their starting points to their final destinations.
    \item Average queue length (AQL): The average queue length is defined as the average number of queuing vehicles waiting in the road network system.
    \item Average waiting time (AWT): The average waiting time measures the average queuing time of vehicles in line at each intersection in the road system
\end{itemize}

\begin{table*}[ht!]
\caption{\uppercase{overall performances of BCT-APLight and previous traditional methods on Jinan and Hangzhou datasets. The best results are highlighted through boldface.}}
\centering
\resizebox{\textwidth}{!}{%
\begin{tabular}{c|ccc|ccc|ccc|ccc|ccc}
\specialrule{0.3mm}{0pt}{0pt} 
\multirow{3}{*}{Method} & \multicolumn{9}{c|}{\textbf{Jinan}}                                                                 & \multicolumn{6}{c}{\textbf{Hangzhou}}                  \\ \cline{2-16} 
                        & \multicolumn{3}{c|}{Dataset-1}     & \multicolumn{3}{c|}{Dataset-2}     & \multicolumn{3}{c|}{Dataset-3}     & \multicolumn{3}{c|}{Dataset-1}     & \multicolumn{3}{c}{Dataset-2}     \\ \cline{2-16} 
                        & ATT   & AQL   & AWT   & ATT   & AQL   & AWT   & ATT   & AQL   & AWT   & ATT   & AQL   & AWT   & ATT   & AQL   & AWT   \\ 
\specialrule{0.2mm}{0pt}{0pt}
Random                  & 604.32 & 693.25 & 101.46  & 565.03 & 434.52 & 103.42 & 622.04 & 292.18 & 95.16  & 632.41 & 325.86 & 75.19  & 601.35 & 689.71 & 94.55  \\ 
FixedTime               & 464.35 & 471.23 & 78.05  & 412.03 & 275.33 & 65.04  & 432.23 & 385.31 & 68.03  & 513.04 & 187.92  & 63.64  & 420.65 & 396.73 & 68.97  \\ 
MaxPressure             & 298.03 & 191.61 & 48.86  & 280.21 & 114.09 & 41.39  & 279.22 & 149.24  &43.77  & 319.82 & 67.52  & 61.82 & 328.59 & 173.97 & 68.82  \\ 
\specialrule{0.2mm}{0pt}{0pt}
MPLight                 & 303.19 & 210.45 & 95.23  & 301.02 & 137.04 & 87.67  & 289.65 & 167.92  & 86.58  & 353.28 & 87.21  & 85.23  & 362.67 & 241.60 & 103.51 \\ 
AttendLight             & 292.35 & 184.33 & 63.21  & 282.84 & 116.92 & 54.28  & 271.03 & 142.87  & 54.68  & 320.54 & 65.29  & 60.92  & 355.23 & 230.47 & 70.61  \\ 
PressLight              & 294.96 & 189.58 & 46.20 & 284.52 & 117.69 & 42.09  & 277.89 & 146.72  & 42.09  & 349.08 & 84.61  & 49.53  & 361.98 & 233.76 & 79.90  \\ 
CoLight                 & 280.26 & 167.41& 57.05  & 271.52 & 105.89 & 53.85  & 264.17 & 130.08  & 48.99  & 315.24 & 67.73  & 60.87  & 332.02 & 189.16 & 86.43  \\ 
Efficient-CoLight       & 276.48 & 176.27 & 47.41  & 269.93 & 102.94 & 39.51  & 264.27 & 130.43 & 42.37  & 308.49 & 55.33  & 32.15  & 337.03 & 185.85 & 67.63  \\ 
Advanced-CoLight        & 274.27 & 157.36 & 48.40  & 266.82 & 100.29 & 43.51  & 262.38 & 128.32  & 43.04  & 300.90 & 48.46 & 39.27  & 326.61 & 168.53 & 73.55  \\ 
\specialrule{0.2mm}{0pt}{0pt}
LightGPT                & 287.60 & 176.17 & 53.21  & 282.24 & 114.93  & 48.78  & 271.34 & 139.27  & 49.02  & 326.40 & 71.55 & 56.92  & 343.42 & 200.43  & 74.99  \\ 
\specialrule{0.2mm}{0pt}{0pt}
BCT-APLight                & \textbf{267.83} & \textbf{147.88} & \textbf{43.34} & \textbf{260.83} & \textbf{93.19} & \textbf{37.28} & \textbf{253.95} & \textbf{117.74} & \textbf{38.90} & \textbf{292.84} & \textbf{42.77} & \textbf{30.65} & \textbf{312.52} & \textbf{143.48} & \textbf{59.74}  \\ 
\specialrule{0.3mm}{0pt}{0pt}
\end{tabular}%
}
\label{tab:comparison}
\end{table*}

\subsection{Comparison experiments}
\label{sec:subsecVB}
This paper evaluates the proposed methods using five real-world traffic flow datasets from Jinan and Hangzhou. Table \ref{tab:comparison} presents a comparative analysis of BCT-APLight and existing methods. 

Among all existing methods, transportation methods generally underperform compared to RL-based methods. In transportation methods, MaxPressure exhibits relatively strong performance. BCT-APLight average decreases \(7.83 \%\) in ATT, \(21.74 \%\) in AQL, and \(20.69\%\) in AWT, respectively. In traditional RL-based methods, Advanced-CoLight achieves the best results; BCT-APLight averages decrease \(3.01 \%\) in ATT, \(9.60 \%\) in AQL, and \(15.28\%\) in AWT, respectively. 

Although LightGPT is built upon the Large Language Model (LLM) framework, its performance remains suboptimal for traffic signal control. Specifically, BCT-APLight achieves an average decrease of \(8.08\%\) in ATT, \(27.94\%\) in AQL, and \(25.81\%\) in AWT compared to LightGPT, demonstrating its superior efficiency. Moreover, LightGPT fails to outperform traditional and RL-based methods such as MaxPressure, CoLight, Efficient-CoLight, and Advanced-CoLight. And due to its significant computational requirements and high training costs, LightGPT is not currently a practical or efficient option for real-world traffic signal control implementations. In comparison, RL-based methods demonstrates high performance and achieve effectiveness in handling traffic signal control tasks. In particular, BCT-APLight consistently achieves state-of-the-art (SOTA) across all baselines. This underscores its exceptional effectiveness and efficiency in addressing the challenges of traffic signal control.

\begin{figure*}[t]
    \centering
    \subfloat[Jinan dataset-1]{%
        \includegraphics[height=7cm]{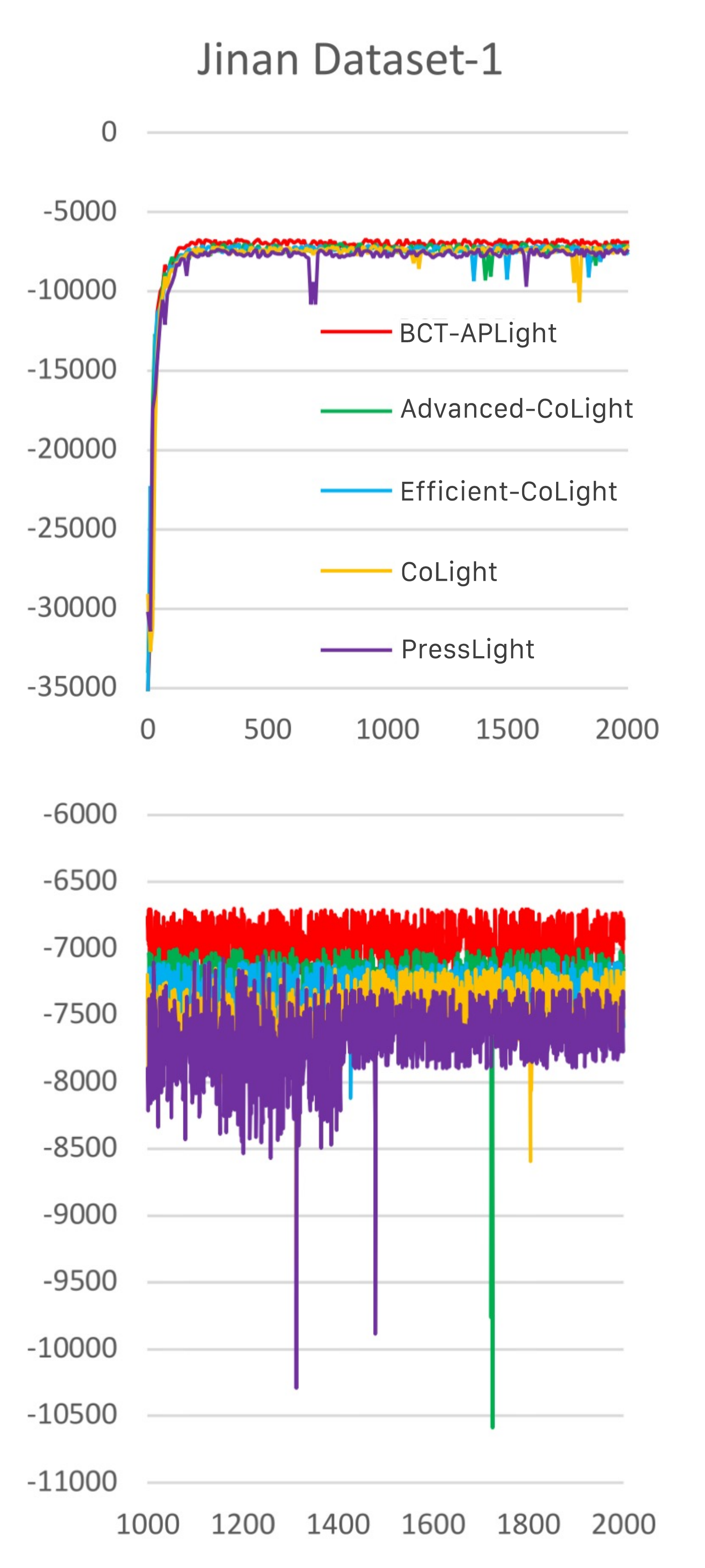}}
    \subfloat[Jinan dataset-2]{%
        \includegraphics[height=7cm]{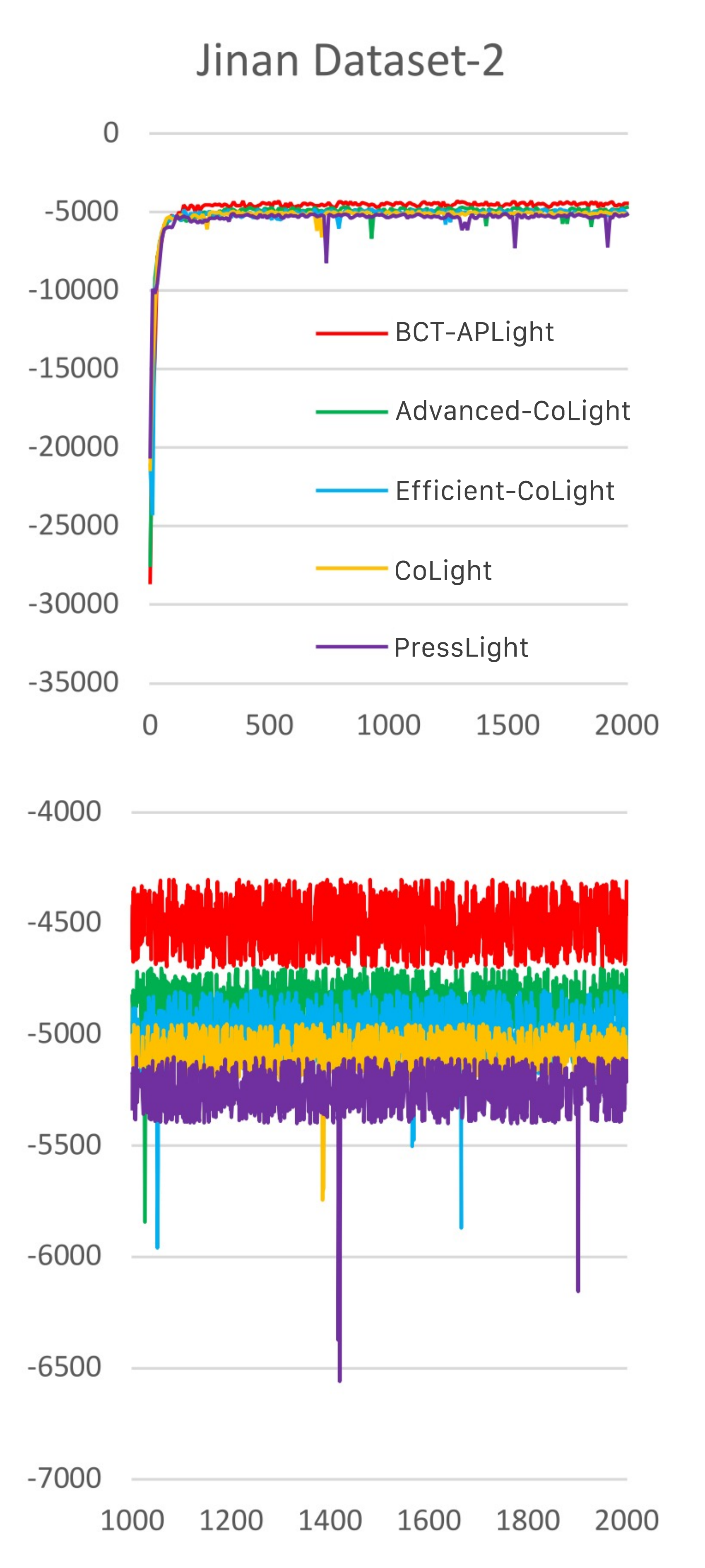}}
    \subfloat[Jinan dataset-3]{%
        \includegraphics[height=7cm]{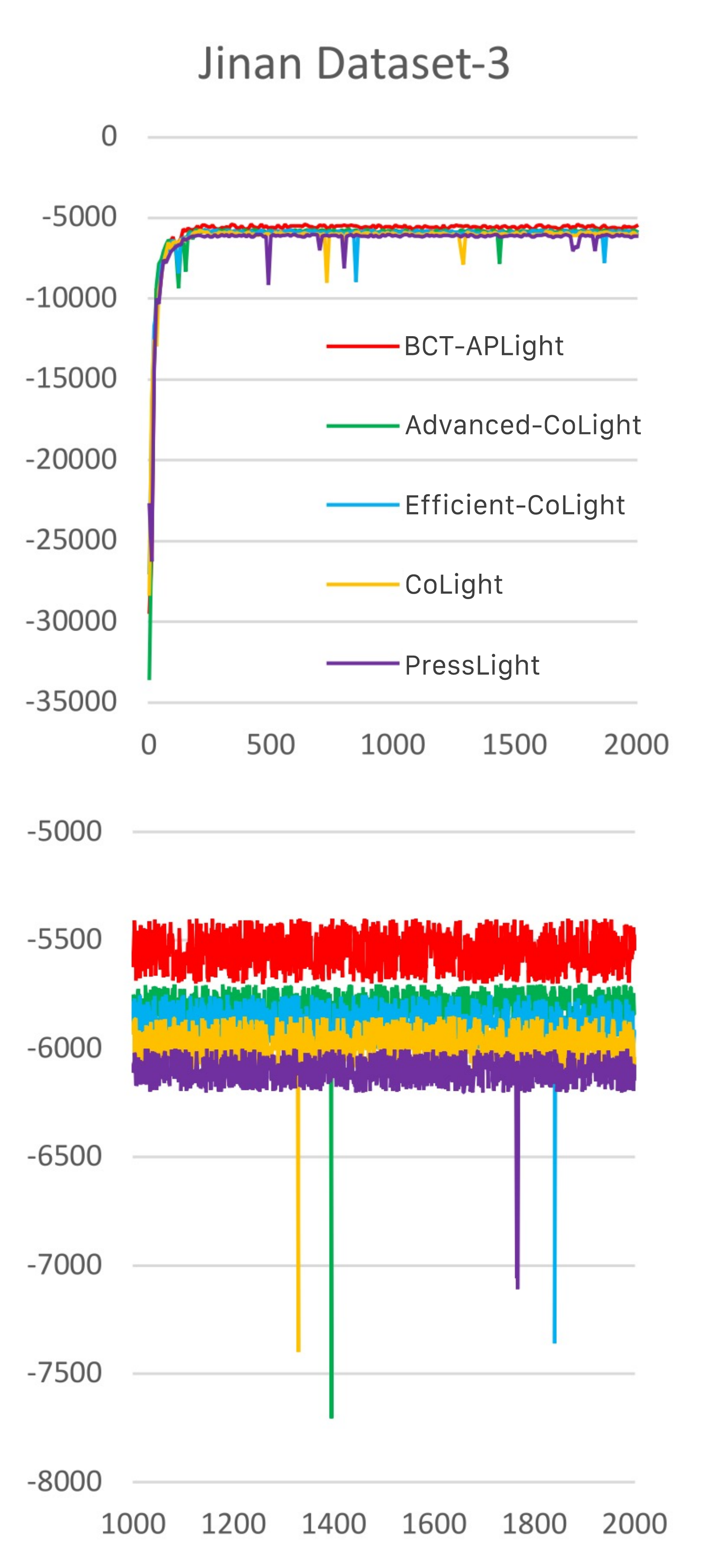}}
    \subfloat[Hangzhou dataset-1]{%
        \includegraphics[height=7cm]{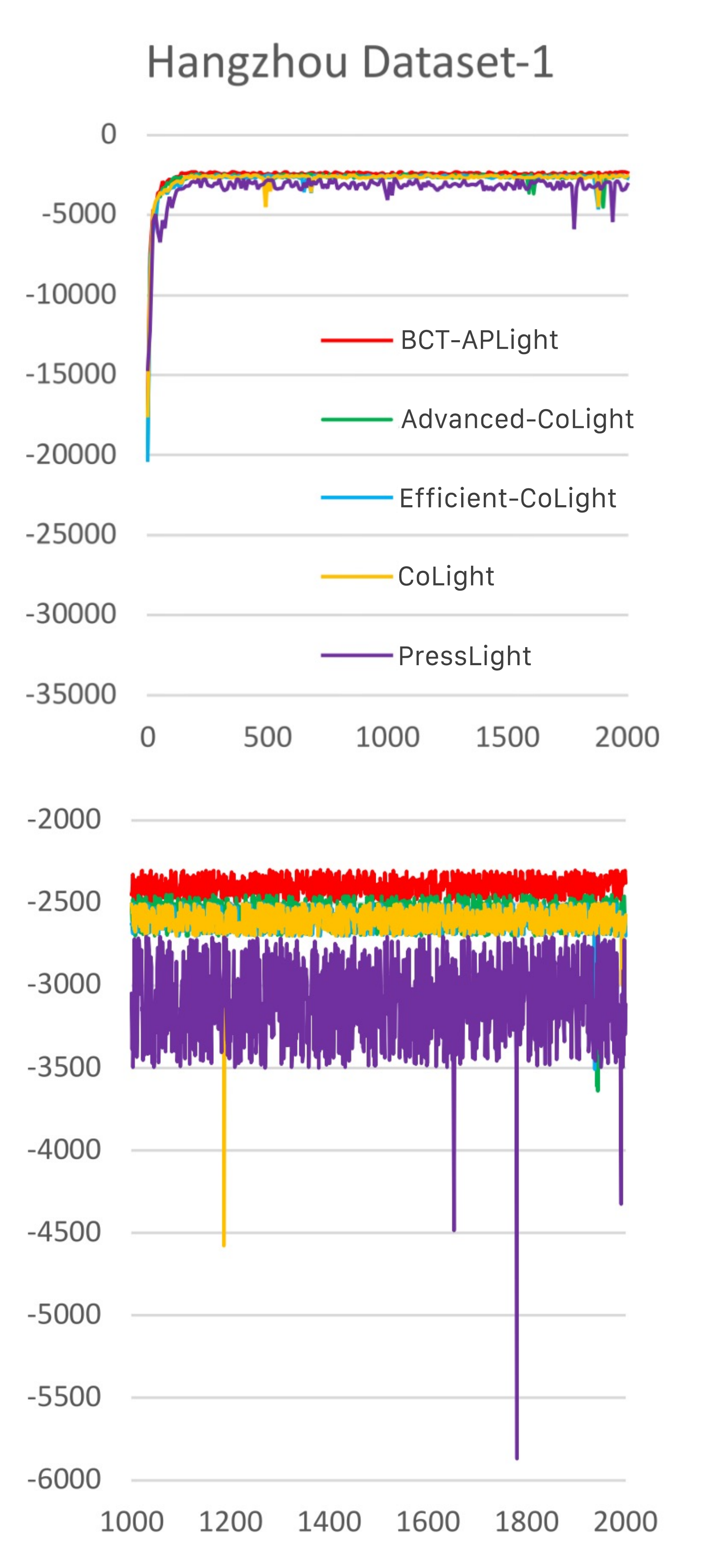}}
    \subfloat[Hangzhou dataset-2]{%
        \includegraphics[height=7cm]{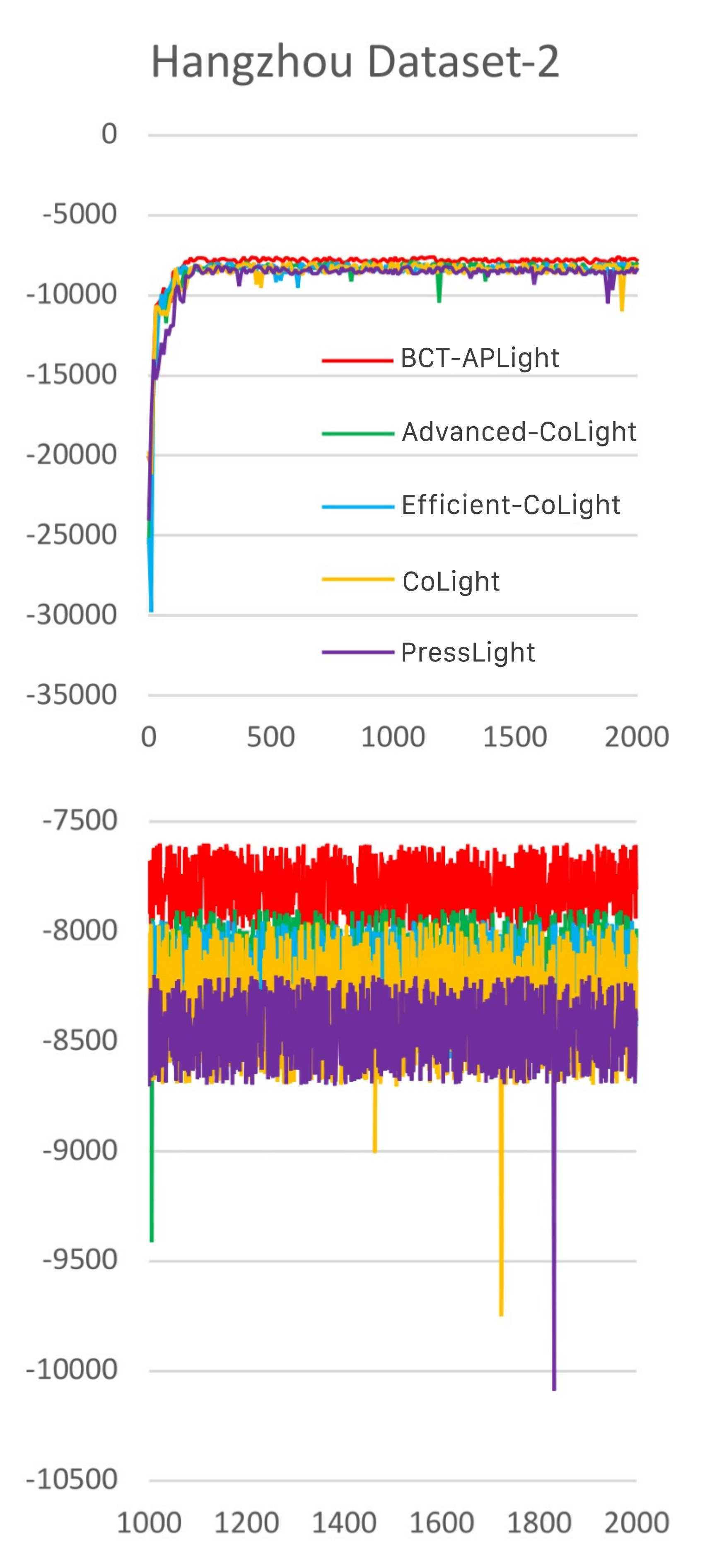}}
    \caption{The twin line chart of the rewards. The upper section illustrates the overall trends of each method over 2000 episodes, while the lower section demonstrates the subtle differences across 1000 episodes after convergence.}
    \label{fig:6}
\end{figure*}

\begin{figure}[ht]
    \centering
    \subfloat[Jinan datasets]{\includegraphics[width=\linewidth,height=3.5cm]{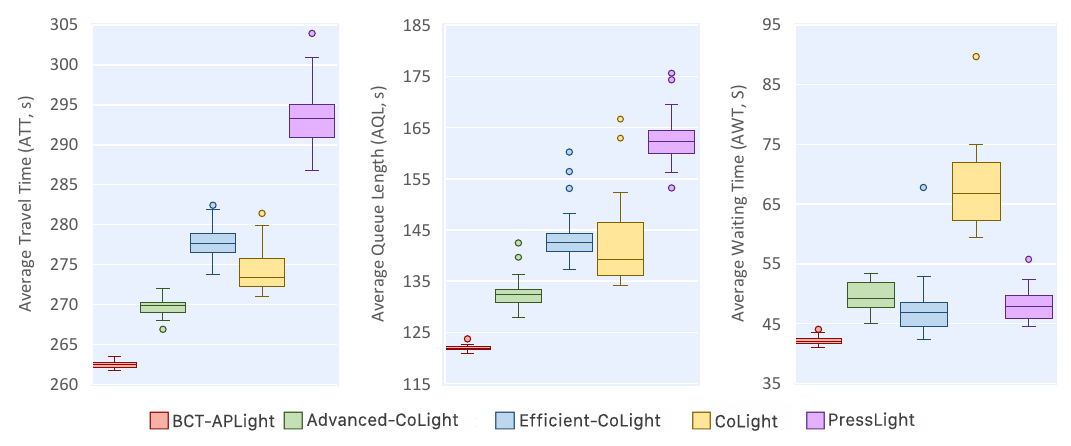}} \\
    \vspace{-0.2cm}
    \subfloat[Hangzhou datasets]{\includegraphics[width=\linewidth,height=3.5cm]{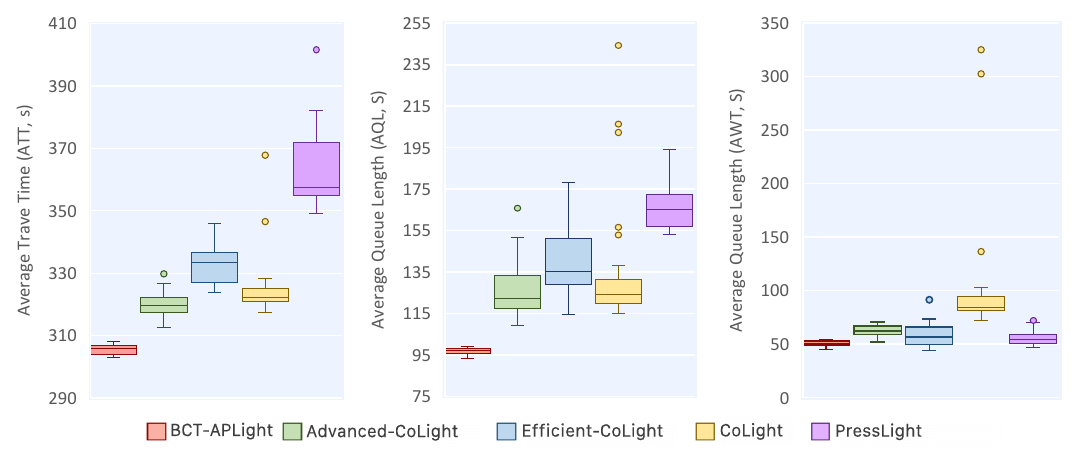}}
    \caption{The comparative results obtained using only the CT framework.}
    \label{fig:7}
\end{figure}

\subsection{Convergence and Stability Analysis} 
\label{sec:subsecVC}
To comprehensively assess the  Convergence and Stability of BCT-APLight, this paper evaluates the train process of BCT-APLight alongside four high-performing traditional RL methods: Advanced-CoLight, Efficient-CoLight, CoLight, and PressLight. 

As shown in Fig. \ref{fig:6}, five twin line charts show the reward trends for each dataset in Jinan and Hangzhou. The upper section illustrates the overall performance of each method over 2000 episodes, capturing the broader training progress. All methods demonstrate the ability to converge effectively during training. The lower section focuses on a smaller reward range and highlights subtle differences across 1000 episodes after convergence. This combined view provides a clear comparison of both the general trends and the finer distinctions between methods. Among all methods, BCT-APLight consistently achieves the most superior performance across training, which highlights its robustness and effectiveness compared to other methods.

Fig. \ref{fig:7} provides two three-group box plots to visualize the average data distribution of ATT, AQL, and AWT in the Jinan and Hangzhou datasets. In detail, the box plot is constructed based on the five-number summary of the data, which includes the minimum, first quartile (Q1), median, third quartile (Q3), and maximum values. The interquartile range (IQR) is calculated as \(IQR = Q3 - Q1\) to determine the lower and upper limits \(Q1 - 1.5 \times IQR\) and \(Q3 + 1.5 \times IQR\) to identify potential outliers. Data points beyond these bounds are considered outliers, while the whiskers extend to the nearest data points within the range. Among the evaluated methods, BCT-APLight demonstrated superior performance consistently across all metrics. This superiority is indicated by its smaller median values and a narrower range in the distribution. In particular, all existing methods exhibit some outliers, whereas BCT-APLight only presents seldom outliers, and the deviation is much smaller. This phenomenon indicates the ability of BCT-APLight to make effective and reasonable policies across all intersections.

\subsection{Large-Scale Intersections Experiments} 
\label{sec:subsecVD}
\begin{figure}[ht]
    \centering
    \subfloat[New York dataset-1]{\hspace{-0.5cm}\includegraphics[width=\linewidth,height=2.8cm]{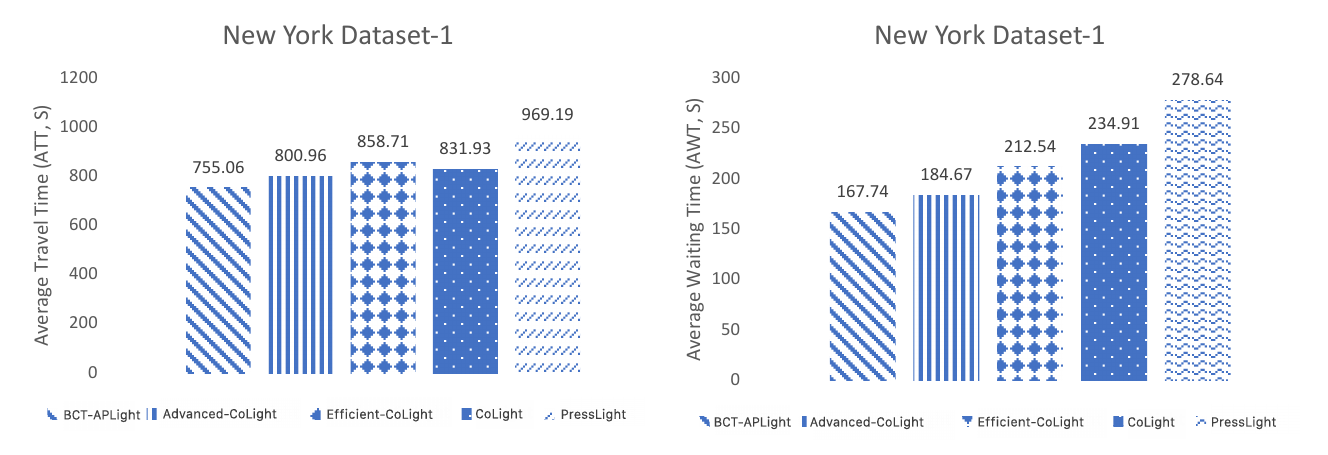}}
    \\
    \vspace{-0.2cm}
    \subfloat[New York dataset-2]{\hspace{-0.5cm}\includegraphics[width=\linewidth,height=2.8cm]{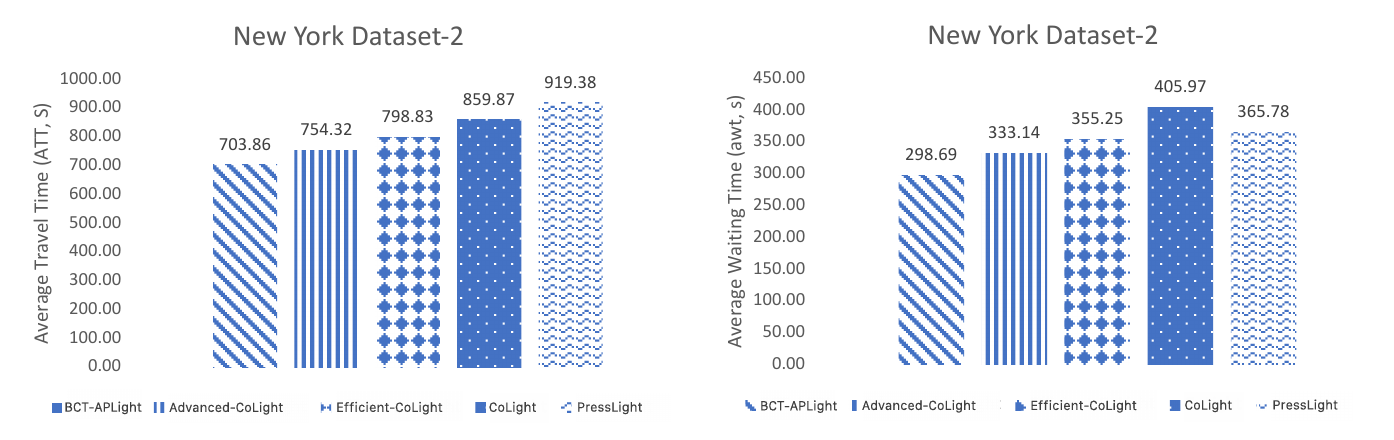}}
    \caption{The comparative results obtained using only the CT framework.}
    \label{fig:8}
\end{figure}

This paper conducts the large-scale intersections experiments of traditional methods and BCT-APLight, evaluating their performance on a significantly larger road network, New York. Fig. \ref{fig:8} provides two two-group bar charts to visualize the data distribution of ATT and AWT in the New York datasets. The results indicate that BCT-APLight achieves the shortest travel and waiting times, Advanced-CoLight demonstrates the best performance among all traditional RL-based methods. Specifically, BCT-APLight achieves a decrease of \(5.73 \%\) in ATT, \(9.17 \%\) in AWT on the New York dataset-1, and \(7.09 \%\) in ATT, \(10.34 \%\) in AWT on the New York dataset-2. Underscoring the exceptional applicability of BCT-APLight to substantially large road networks.

From the charts, it is evident that as the road network complexity increases from New York dataset-1 to dataset-2, all methods experience an increase in both ATT and AWT. However, the relative performance gap between BCT-APLight and other methods remains significant. For example, in New York dataset-1, the ATT of BCT-APLight (755.06s) is 45.90s lower than Advanced-CoLight and 214.13 s lower than PressLight. Similarly, for dataset-2, BCT-APLight outperforms Advanced-CoLight by 50.46s and PressLight by an even larger margin of 215.52s. A similar trend is observed for AWT, with BCT-APLight achieving substantial improvements in both datasets, indicating its robustness and scalability for managing traffic in large-scale networks effectively. 

\begin{table*}[ht!]
\caption{RESULTS OF ABLATION STUDY}
\centering
\resizebox{\textwidth}{!}{%
\begin{tabular}{c|ccc|ccc|ccc|ccc|ccc}
\specialrule{0.3mm}{0pt}{0pt} % 第一行粗横线
\multirow{3}{*}{Method} & \multicolumn{9}{c|}{\textbf{Jinan}}                                                                 & \multicolumn{6}{c}{\textbf{Hangzhou}}                  \\ \cline{2-16} 
                        & \multicolumn{3}{c|}{Dataset-1}     & \multicolumn{3}{c|}{Dataset-2}     & \multicolumn{3}{c|}{Dataset-3}     & \multicolumn{3}{c|}{Dataset-1}     & \multicolumn{3}{c}{Dataset-2}     \\ \cline{2-16} 
                        & ATT   & AQL   & AWT   & ATT   & AQL   & AWT   & ATT   & AQL   & AWT   & ATT   & AQL   & AWT   & ATT   & AQL   & AWT   \\ 
\specialrule{0.2mm}{0pt}{0pt}
DQN                  & 368.43 & 270.35 & 100.06  & 345.71 & 205.93 & 106.42 & 327.11 & 249.68 & 96.06  & 427.65 & 143.26 & 93.99  & 440.62 & 307.35 & 143.61 \\ 
\textbf{AP-Based DQN}            & 268.45 & 151.00 & 47.54  & 263.75 & 96.87 & 40.16  & 255.61 & 119.90  & 41.21  & 295.83 & 44.30 & 36.29  & 317.62 & 148.49 & 60.52 \\ 
\specialrule{0.2mm}{0pt}{0pt}
Advanced-CoLight        & 274.27 & 157.36 & 48.40  & 266.82 & 100.29 & 43.51  & 262.38 & 128.32  & 43.04  & 300.90 & 48.46 & 39.27  & 326.61 & 168.53 & 73.55  \\ 
\textbf{CT-Based Advanced-CoLight}  & 270.53 & 152.67 & 45.65  & 264.16 & 97.58 & 40.04  & 257.09 & 121.10  & 42.05  & 297.00 & 45.34 & 33.26  & 317.06 & 157.13 & 69.41  \\ 
\specialrule{0.2mm}{0pt}{0pt}
BCT-APLight                & \textbf{267.83} & \textbf{147.88} & \textbf{43.34} & \textbf{260.83} & \textbf{93.19} & \textbf{37.28} & \textbf{253.95} & \textbf{117.74} & \textbf{38.90} & \textbf{292.84} & \textbf{42.77} & \textbf{30.65} & \textbf{312.52} & \textbf{143.48} & \textbf{59.74}\\
\specialrule{0.3mm}{0pt}{0pt} % 最后一行粗横线
\end{tabular}%
}
\label{tab:ablation study}
\end{table*}

\subsection{Ablation Study}
\label{sec:subsecVE}
This paper evaluates the performance of each component in BCT-APLight, namely the CT framework and the AP mechanism. Firstly, the effectiveness of the AP mechanism is examined by comparing the original DQN algorithm with the AP-based DQN. As shown in Fig. \ref{tab:ablation study}, the AP-based DQN significantly outperforms the original DQN algorithm. Specifically, the AP mechanism leads to substantial reductions of \(26.62 \%\) in ATT, \(52.36 \% \) in AQL, and \(58.21 \%\) in AWT. These results highlight the ability of the AP mechanism to optimize the traffic movement representation, thereby substantially improving traffic signal control efficiency.

Additionally, this paper explores the effectiveness of the CT framework by incorporating it into the convergence model of Advanced-CoLight, the best-performing traditional RL-based method, for further training. The results show that integrating the CT framework yields improvements of \(1.74 \%\) in ATT, \(4.83 \%\) in AQL, and \(7.01 \%\) in AWT over Advanced-CoLight. While the relative improvement is smaller compared to the AP-based DQN experiment, this is expected because Advanced-CoLight already achieves near-optimal performance among traditional methods. Nevertheless, the fact that the CT framework can still deliver measurable gains on such a strong baseline underscores its robustness and effectiveness in refining RL-based methods. These results highlight the versatility and value of the CT framework, even when applied to well-optimized models.

Moreover, even when only the AP mechanism is applied without the CT framework, the performance of surpasses all traditional methods, including Advanced-CoLight. This finding underscores the independent effectiveness of the AP mechanism. However, the removal of either the AP mechanism or the CT framework results in a marked degradation of the performance of model, reaffirming the critical role of both components. In conclusion, this ablation study underscores the importance of the CT framework and AP mechanism in advancing the capabilities of RL-based methods for ATSC.

\section{Conclusion}
\label{sectionVI}

The Adaptive Traffic Signal Control (ATSC) system plays a vital role in intelligent transportation, offering the potential to effectively mitigate urban traffic congestion. While reinforcement learning (RL)-based methods have shown great promise in enhancing ATSC, current methods struggle with generating rational and effective policies. In this paper, a novel RL-based method (called BCT-APLight) is proposed for optimizing multi-intersection TSC. The BCT-APLight integrates two components: the Critique-Tune (CT) framework and the Attention-Based Adaptive Pressure (AP). Among it, the Critique Layer evaluate the credibility of policies, while the Tune Layer minimizes posterior risks to fine-tune policies when the evaluation is negative. This hierarchical mechanism ensures a robust policy refinement process, enabling more effective decision-making under varying traffic conditions. Meanwhile, the AP mechanism introduces an attention-based adaptive pressure to effectively weight the vehicle queues in each lane, thereby providing a precise and dynamic representation of traffic movement. Experiments conducted across seven real-world datasets with diverse intersection layouts validate the superiority of BCT-APLight over existing state-of-the-art methods. The results demonstrate that BCT-APLight achieves improved traffic flow efficiency, enhancing the reasonableness of RL policies in real-world traffic scenarios.

In the future, the predictive accuracy of the prediction network can be further enhanced by incorporating more advanced modeling techniques or integrating additional real-time traffic data. Furthermore, the transferability of BCT-APLight across diverse traffic scenarios and network layouts is a promising avenue for exploration, ensuring its adaptability and scalability to address future challenges in transportation efficiency and smart city development.

% or

% \subsection or using \label (\appendices by itself\(\(\(\(\)\)\)\)

\appendices
\section{Theoretical Explanation of The Identical ARIMA Models in SARIMA}\label{proof 1}
According to Eq. \ref{equation3} and Eq. \ref{equation7}, applying the zero-mean transformation to the differenced series \(w_t^{i(d)}\):
\begin{equation}
y_{n,i(d)} = w_t^{n,i(d)} - \bar{w}_t^{i(d)},
\end{equation}
where \(\bar{w}_t^{i(d)}\) is the mean of the \(n-th\) episode differenced series.

Given that the mean does not affect stationarity:
\begin{align}
E[y_t^{n,i(d)}] &= E[w_t^{n,i(d)} - \bar{w}_t^{i(d)}]\nonumber\\
&= E[w_t^{n,i(d)} - \bar{w}_t^{i(d)}] - \bar{w}_t^{i(d)}\nonumber\\
&=0,
\end{align}
the variance and autocorrelation structure thus remain unchanged:
\begin{equation}
\text{Var}(y_t^{n,i(d)} = \text{Var}(w_t^{n,i(d)}),
\end{equation}
\begin{equation}
\quad \text{Cov}(y_t^{n,i(d)}, y_t^{n+s,i(d)}) = \text{Cov}(w_t^{n,i(d)}, w_t^{n+s,i(d)}),
\end{equation}
Therefore, \(y_t^{n,i(d)}\) remains stationary.

Assume that the original differenced series \(w_t^{i(d)}\) follows the ARIMA model with parameters \(\phi_k\) and \(\theta_j\). Substituting \(y_t^{n,i(d)} = w_t^{n,i(d)} - \bar{w}_t^{i(d)}\) into the ARIMA model:
\begin{equation}
y_t^{n,i(d)} + \bar{w}_t^{i(d)} = \sum_{k=1}^p \phi_k (y_t^{n-k,i(d)}+ \bar{w}_t^{i(d)}) + \epsilon_{n,i} + \sum_{j=1}^q \theta_j \epsilon_{n-j,i}.
\end{equation}

Expanding above equation:
\begin{align}
y_t^{n,i(d)} &= \sum_{k=1}^p \phi_k y_t^{n-k,i(d)}+\epsilon_{n,i} + \sum_{j=1}^q \theta_j \epsilon_{n-j,i}\nonumber\\
&+ \sum_{k=1}^p \phi_k \bar{w}_t^{i(d)} - \bar{w}_t^{i(d)}.
\end{align}

The term  \(\sum_{k=1}^p \phi_k \bar{w}_t^{i(d)} - \bar{w}_t^{i(d)}\) simplifies to 0 because the ARIMA coefficients \(\phi_k\) satisfy the normalization property  \(\sum_{k=1}^p \phi_k = 1\). Therefore, the equation reduces to:
\begin{align}
y_t^{n,i(d)} = \sum_{k=1}^p \phi_k y_t^{n-k,i(d)}+\epsilon_{n,i} + \sum_{j=1}^q \theta_j \epsilon_{n-j,i}.
\end{align}

Overall, the aforementioned process shows that the zero-mean-transformed series \(y_t^{n,i(d)}\) follows the identical ARIMA model with the same parameters \(\phi_k\) and \(\theta_j\).

\section{Proof of the Bayesian posterior risk process}\label{proof 2}
In the Bayesian statistical decision problem, the definition of risk function is expressed as follows:
\begin{equation}
R(\delta, \theta) = E[L(\delta, \theta)] = \int_{\mathcal{X}} L(\delta, \theta) f(x, \theta) dx,
\end{equation}
where \textit{$\delta$} is the decision rule, and the average loss \(R(\delta, \theta)\) is defined as the risk function of \(\delta\).

According to Wald’s statistical decision theory, the only criterion for evaluating the decision rules \(\delta\) is the associated risk function. If there exists a decision rule \textit{$\delta^*$} such that for all \textit{$\theta \in \Theta$} satisfy \(R(\delta^*, \theta) \leq R(\delta, \theta)\), the \textit{$\delta^*$} is called an admissible decision rule or a uniformly better decision rule. In practice, such a decision rule is often absent. As a result, the criteria are always relaxed; the common method is to impose optimality conditions based on the Bayes criterion.

Given a risk function \(R(\delta, \theta)\) and a prior distribution \(H(\theta)\) on \(\theta\), with the prior denoted by \(\pi(\theta)\), the Bayes risk is defined as:
\begin{align}
\label{equation2}
R_H(\delta) &= E^{\theta}[R(\delta, \theta)] = \int_\Theta R(\delta, \theta) dH(\theta) \nonumber \\
&= \int_\Theta \left( \int_{\mathcal{X}} L(\delta, \theta) f(x, \theta) dx \right) dH(\theta),
\end{align}
The decision rule that minimizes the Bayes risk is termed the Bayes solution, denoted as \(R_H(\delta^*)\).

After that, suppose \textit{$\theta$} is fixed, and the random variable \textit{$X$} follows the distribution \textit{$f(x|\theta)$}. When new data is observed, the posterior distribution of \textit{$\theta$} is updated to \textit{$H(\theta|x)$}, using the prior $H(\theta)$ and likelihood \textit{$f(x|\theta)$}. The risk function is now evaluated with respect to the posterior distribution:
\begin{equation}
R(\delta|x) = E^{\theta|x}[L(\delta, \theta)] = \int_\Theta L(\delta, \theta) dH(\theta|x),
\end{equation}
This is called the posterior risk of the decision rule \textit{$\delta$}. If there exists a decision rule \textit{$\delta^*$} such that:
\begin{equation}
R(\delta^* | x) = \min_{\delta} R(\delta | x),
\end{equation}
then \(\delta^*\) is called the Bayes solution that minimizes the posterior risk.

From the Eq. (\ref{equation2}), the following expression holds:
\begin{align}
R_H(\delta) &= E^{\theta}[R(\delta, \theta)] = \int_\Theta R(\delta, \theta) dH(\theta)\nonumber \\
&= \int_\Theta \left[ \int_{\mathcal{X}} L(\delta, \theta) F(x|\theta) \right] dH(\theta) \nonumber \\
&= \int_{\mathcal{X}} \left[ \int_\Theta L(\delta, \theta) d(\theta|x) dx \right]dF_m(x)\nonumber \\
&=\int_{\mathcal{X}} R(\delta | x) dF_m(x) = E^{X}[R(\delta|x)],
\end{align}
where \(F_m(x)\) is the marginal distribution of \textit{X}, and \(f_m(x)\) is its density. \(E^{\theta}\) represents the expectation taken with respect to the prior distribution of \(\theta\). Therefore, the Bayes risk can be written as \(E^X[R(\delta | x)]\) indicating that the Bayes risk is simply the expectation of the posterior risk with respect to the marginal distribution of \textit{$X$}.

It can be shown that the decision rule \textit{$\delta^*$} that minimizes the posterior risk\(R(\delta^* | x)\) is the Bayes solution under the prior distribution \(H(\theta)\). Besides, this Bayes solution \(R_H(\delta^*)\) also minimizes the Bayes risk \textit{$R_H(\delta^*)$}, as shown by the following:
\begin{align}
R(\delta | x) &= \int_\Theta L(\delta, \theta) H(d\theta | x) \nonumber \\
&\geq \int_\Theta L(\delta_H, \theta) H(d\theta | x) = R(\delta_H | x),
\end{align}
By taking the expectation of both sides with respect to the marginal distribution $F_m(x)$, the following inequality is obtained:
\begin{align}
R_H(\delta) &= \int_{\mathcal{X}} R(\delta | x) dF_m(x) \nonumber \\
&\geq \int_{\mathcal{X}} R(\delta_H | x) dF_m(x) = R_H(\delta_H),
\end{align}
Thus, \textit{$\delta_H$} minimizes the Bayes risk and is the Bayes solution, which achieves the minimal risk of the policy-making process of RL in ATSC.

\ifCLASSOPTIONcaptionsoff
  \newpage
\fi



%\bibliographystyle{IEEEtran}


% \end{thebibliography}

\bibliographystyle{IEEEtran}
\bibliography{reference}

% needed aepisode the contents of the optional argument to biography to prevent
\begin{IEEEbiography}[{\includegraphics[width=1in,height=1.25in,clip,keepaspectratio]{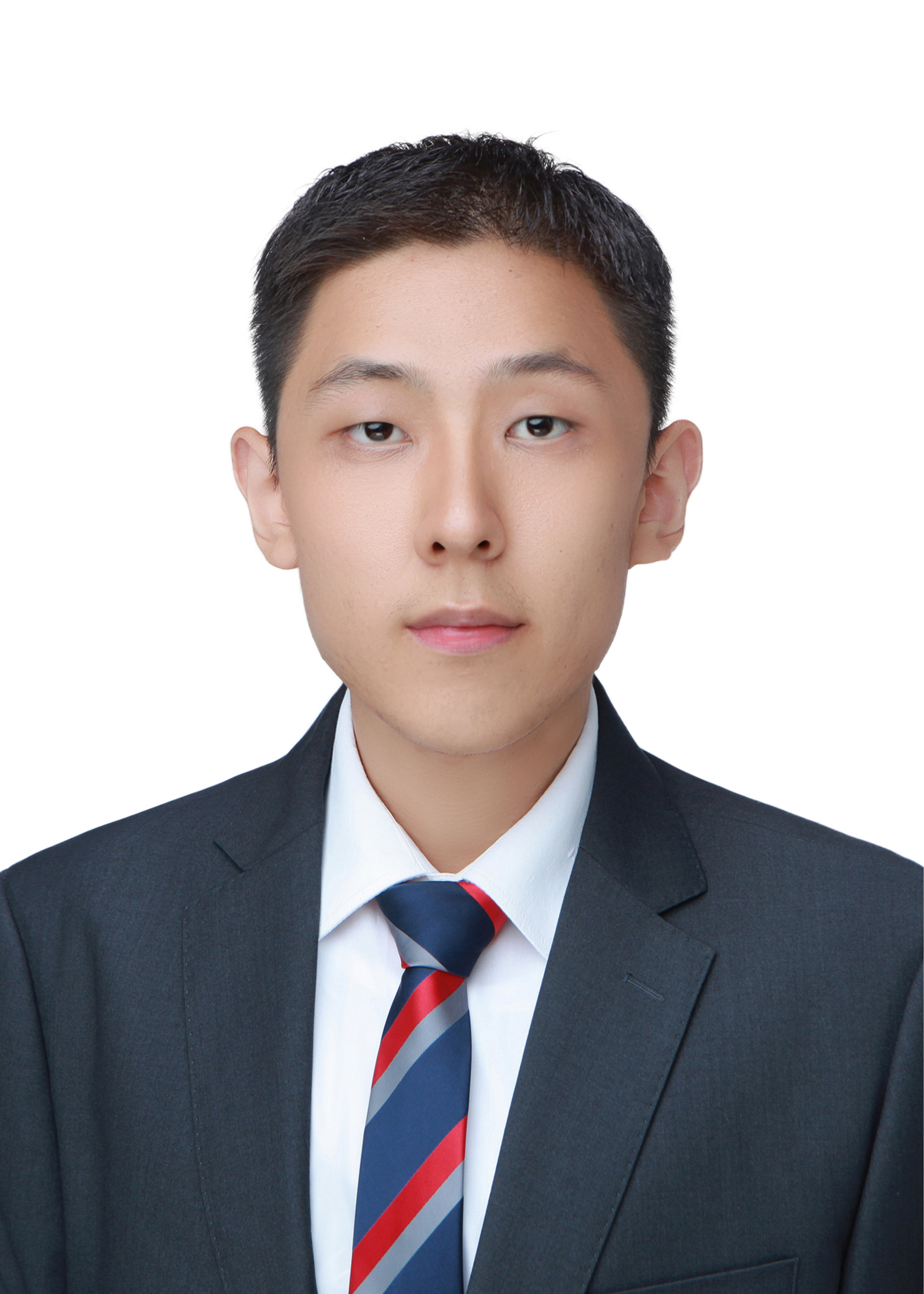}}]{Wenchang Duan} (Graduate Student Member, IEEE)
received the B.S. degree in statistics from the Xiangtan University, Xiangtan, China, where he is currently pursuing the M.S. degree in Applied Statistics with Shanghai Jiao Tong University, Shanghai, China. His research interests include deep reinforcement learning, intelligent transportation systems, autonomous driving systems, and mathematical statistics.
\end{IEEEbiography}

\begin{IEEEbiography}[{\includegraphics[width=1in,height=1.25in,clip,keepaspectratio]{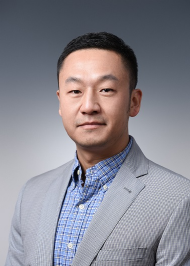}}]{Zhenguo Gao} is an assistant professor in the School of Mathematical Sciences, Shanghai Jiao Tong University in China. He received his PhD degree in statistics from Virginia Tech US in 2018. His research area includes mathematical statistics, deep reinforcement learning, data mining and machine learning, intelligent transportation systems and high-dimensional data analysis.
\end{IEEEbiography}

\begin{IEEEbiography}[{\includegraphics[width=1in,height=1.25in,clip,keepaspectratio]{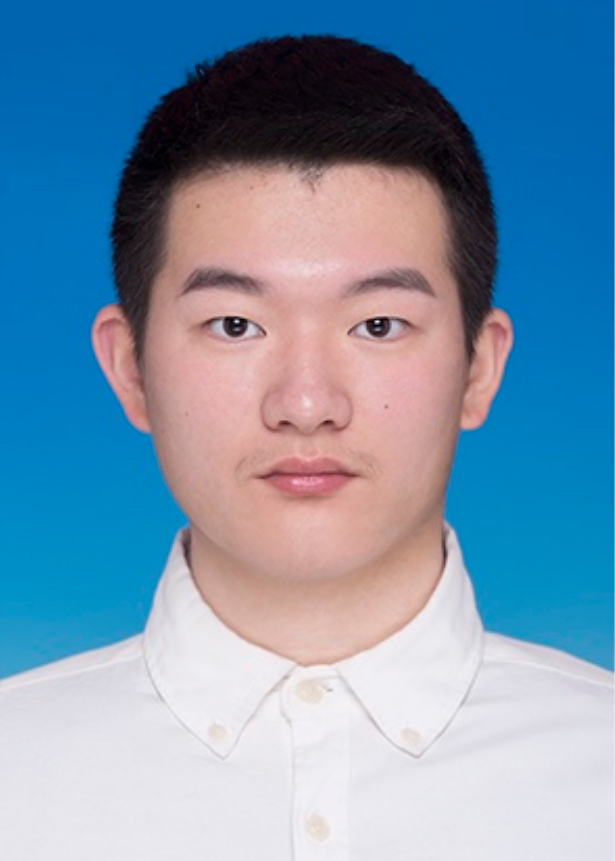}}]{Jiwan He} Jiwan He received the B.S. degree in Information and Computational Science from Xi’an Jiaotong University, Xi’an, China. He is currently pursuing the M.S. degree in Applied Statistics with Shanghai Jiao Tong University, Shanghai, China. His research interests include deep learning, Bayesian statistics, and intelligent transportation systems.
\end{IEEEbiography}

\begin{IEEEbiography}[{\includegraphics[width=1in,height=1.25in,clip,keepaspectratio]{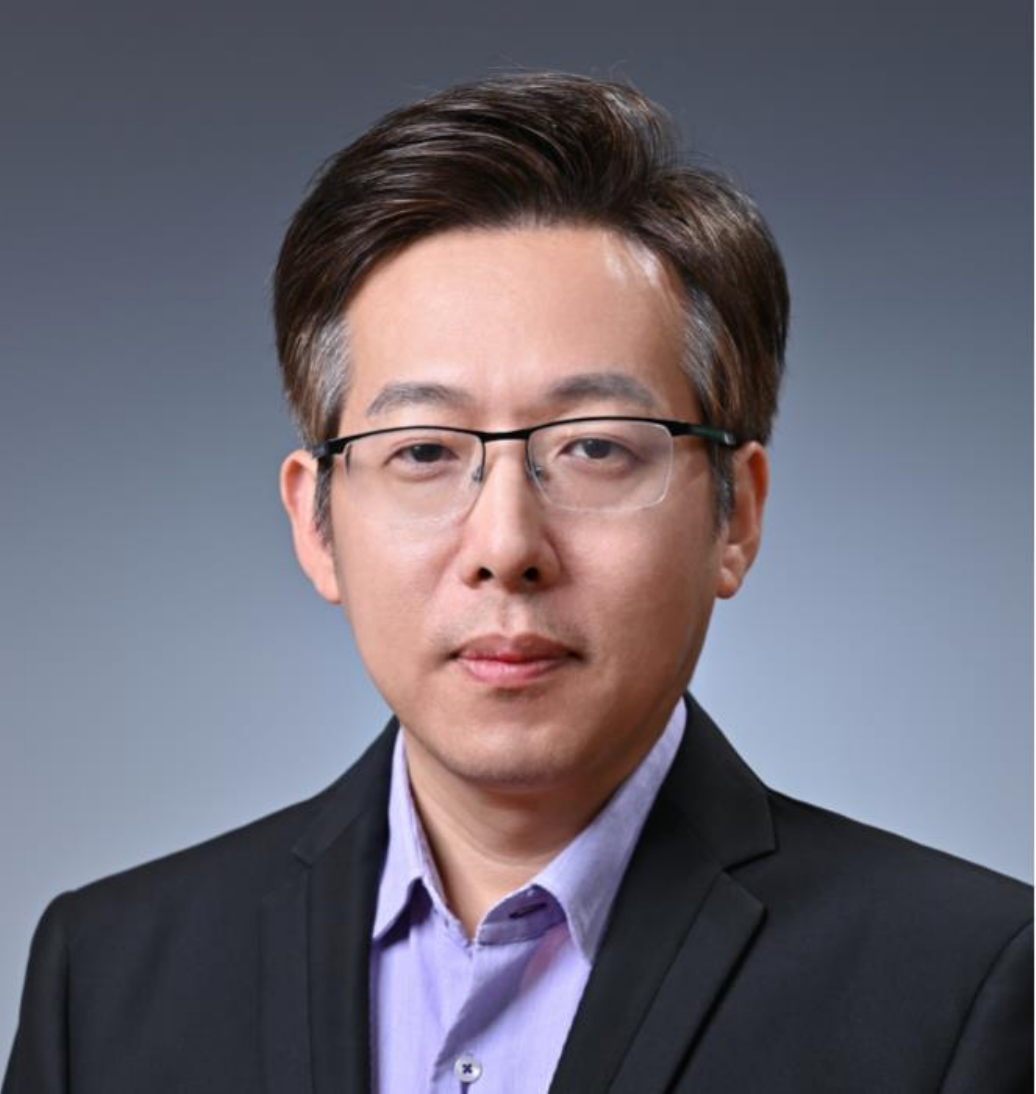}}]{Jinguo Xian} is an associate professor in the School of Mathematical Sciences, Shanghai Jiao Tong University in China. He received his PhD degree in Mathematics from Shanghai Jiao Tong University. His research area includes stochastic process monitoring, change point detection, random network, deep reinforcement learning and intelligent transportation systems.
\end{IEEEbiography}

% \end{IEEEbiography}


% % insert where needed to balance the two columns on the last page with
% %\newpage



%\vfill


\end{document}